\documentclass{article}

\usepackage{PRIMEarxiv}

\PassOptionsToPackage{off}{epstopdf}
\usepackage[off]{epstopdf}

\usepackage{pgfplots}
\usepackage{pgfplotstable}
\usepackage[numbers]{natbib}
\pgfplotsset{compat=1.18}
\usepgfplotslibrary{groupplots}

\usepackage{hyperref}
\usepackage{xr-hyper}

\usepackage[T1]{fontenc}
\usepackage{inconsolata} 
\usepackage{pifont}

\newlength{\ttcharwd}
\AtBeginDocument{\settowidth{\ttcharwd}{\texttt{0}}}
\newcommand{\cmarkbox}{\makebox[\ttcharwd][c]{\ding{51}}}
\newcommand{\xmarkbox}{\makebox[\ttcharwd][c]{\ding{55}}}

\usepackage[utf8]{inputenc} 
\usepackage[T1]{fontenc}    
\usepackage{url}            
\usepackage{booktabs}       
\usepackage{amsfonts}       
\usepackage{nicefrac}       
\usepackage{microtype}      
\usepackage{lipsum}
\usepackage{fancyhdr}       
\usepackage{graphicx}       
\graphicspath{{media/}}     
\usepackage{verbatim}       

\usepackage{titlesec}
\titleformat{\paragraph}[runin]{\bfseries}{}{}{}[.]
\titlespacing*{\paragraph}{0pt}{0.7\baselineskip}{0.5em}

\usepackage{amsmath}
\usepackage{amsfonts}

\usepackage{tikz}
\usetikzlibrary{
  arrows.meta,
  positioning,
  calc,
  shadows.blur,
  fit,
  matrix,
  backgrounds
}
\usetikzlibrary{shapes.geometric, arrows, positioning}

\usepackage[british]{babel}
\usepackage{hhline}
\usepackage{multirow}
\usepackage{subcaption}
\usepackage{colortbl}
\usepackage{makecell} 
\usepackage{colortbl} 
\usepackage{makecell} 
\usepackage{amsmath} 
\usepackage{array}
\usepackage{caption}  

\usepackage{float}  

\makeatletter
\newcommand{\beginsupplement}{%
  \setcounter{section}{0}
  \setcounter{subsection}{0}
  \setcounter{subsubsection}{0}
  \setcounter{figure}{0}
  \setcounter{table}{0}
  \setcounter{equation}{0}
  \setcounter{lstlisting}{0}
  \setcounter{footnote}{0}
  \setcounter{Hfootnote}{0}

  \renewcommand{\thesection}{S\arabic{section}}
  \renewcommand{\thesubsection}{S\arabic{section}.\arabic{subsection}}
  \renewcommand{\thesubsubsection}{S\arabic{section}.\arabic{subsection}.\arabic{subsubsection}}
  \renewcommand{\thefigure}{S\arabic{figure}}
  \renewcommand{\thetable}{S\arabic{table}}
  \renewcommand{\theequation}{S\arabic{equation}}
  \renewcommand{\thelstlisting}{S\arabic{lstlisting}}
  \renewcommand{\thefootnote}{S\arabic{footnote}}

  \renewcommand{\theHsection}{S\arabic{section}}
  \renewcommand{\theHsubsection}{S\arabic{section}.\arabic{subsection}}
  \renewcommand{\theHsubsubsection}{S\arabic{section}.\arabic{subsection}.\arabic{subsubsection}}
  \renewcommand{\theHfigure}{S\arabic{figure}}
  \renewcommand{\theHtable}{S\arabic{table}}
  \renewcommand{\theHequation}{S\arabic{equation}}
  \renewcommand{\theHlstlisting}{S\arabic{lstlisting}}
  \renewcommand{\theHfootnote}{S\arabic{footnote}}
}
\makeatother

\usepackage{booktabs}
\usepackage{siunitx}
\usepackage{threeparttable}

\usepackage{listings}
\usepackage{xcolor}

\definecolor{pythonblue}{RGB}{32,74,135}
\definecolor{pythonstring}{RGB}{164,0,0}
\definecolor{pythoncomment}{RGB}{96,96,96}

\lstdefinestyle{pythonstyle}{
    language=Python,
    basicstyle=\ttfamily\small,
    commentstyle=\color{pythoncomment},
    stringstyle=\color{pythonstring},
    keywordstyle=\color{pythonblue},
    showstringspaces=false,
    numbers=left,
    numberstyle=\tiny\color{pythoncomment},
    frame=single,
    rulecolor=\color{black},
    backgroundcolor=\color{white},
    breaklines=true,
    breakatwhitespace=true,
    tabsize=4
}

\definecolor{backcolour}{rgb}{0.95,0.95,0.92}
\definecolor{commentcolour}{rgb}{0.5,0.5,0.5}
\definecolor{stringcolour}{rgb}{0.58,0,0.82}
\definecolor{keywordcolour}{rgb}{0,0,0.6}
\definecolor{numbercolour}{rgb}{0.1,0.1,0.5}

\definecolor{oiBlue}{RGB}{0,114,178}        
\definecolor{oiOrange}{RGB}{230,159,0}      
\definecolor{oiSkyBlue}{RGB}{86,180,233}    
\definecolor{oiBluishGreen}{RGB}{0,158,115} 

\definecolor{emerald}{RGB}{16,185,129}
\definecolor{crimson}{RGB}{220,38,127}
\definecolor{slate}{RGB}{100,116,139}

\lstdefinelanguage{json}{
    basicstyle=\ttfamily\small,
    keywordstyle=\color{blue},
    stringstyle=\color{purple},
    commentstyle=\color{gray},
    morestring=[b]",
    morecomment=[l]{//},
    morecomment=[s]{/*}{*/},
    showstringspaces=false,
    showspaces=false,
    showtabs=false,
    tabsize=2,
    breaklines=true,
    frame=single
}

\definecolor{backcolour}{rgb}{0.95,0.95,0.92}
\definecolor{commentcolour}{rgb}{0.5,0.5,0.5}
\definecolor{stringcolour}{rgb}{0.58,0,0.82}
\definecolor{keywordcolour}{rgb}{0,0,0}
\definecolor{numbercolour}{rgb}{0.1,0.1,0.5}

\title{EnvTrace: Simulation-Based Semantic Evaluation of LLM Code via Execution Trace Alignment-- Demonstrated at Synchrotron Beamlines}

\author{
  Noah van der Vleuten\\
  Center for Functional Nanomaterials\\
  Brookhaven National Laboratory \\
  Upton, NY 11973, USA\\
  \And
  Anthony Flores \\  
  Center for Functional Nanomaterials\\
  Brookhaven National Laboratory \\
  Upton, NY 11973, USA\\
  \And
  Shray Mathur \\
  Center for Functional Nanomaterials\\
  Brookhaven National Laboratory \\
  Upton, NY 11973, USA\\
  \And
  Max Rakitin \\  
  National Synchrotron Light Source II\\
  Brookhaven National Laboratory \\
  Upton, NY 11973, USA\\
  \And
  Thomas Hopkins \\ 
  National Synchrotron Light Source II\\
  Brookhaven National Laboratory \\
  Upton, NY 11973, USA\\
  \And
  Kevin G. Yager \\
  Center for Functional Nanomaterials\\
  Brookhaven National Laboratory \\
  Upton, NY 11973, USA\\
  \And
  Esther Tsai\\
  Center for Functional Nanomaterials\\
  Brookhaven National Laboratory \\
  Upton, NY 11973, USA\\
  \texttt{etsai@bnl.gov} \\
}

\begin{document}
\maketitle

\begin{abstract}
Evaluating large language models (LLMs) for instrument control requires methods that go beyond standard, stateless algorithmic benchmarks, since the behavior of physical systems cannot be fully captured by unit tests alone.
Here we introduce EnvTrace, a simulation-based method that evaluates execution traces to assess semantic code equivalence. 
EnvTrace is demonstrated with a beamline control-logic digital twin to facilitate the evaluation of instrument control code, with the digital twin itself also 
enabling the pre-execution validation of live experiments. Over 30 LLMs were evaluated using trace alignment to generate a multi-faceted score for functional correctness across key behavioral dimensions, showing that many top-tier models can approach human-level performance in rapid control-code generation.
This is a first step toward a broader vision where LLMs and digital twins work symbiotically: LLMs providing intuitive control and agentic orchestration, and digital twins offering safe and high-fidelity environments, paving the way towards autonomous embodied AI.

\end{abstract}

\section{Introduction}

The advent of large language models (LLMs) has catalyzed a paradigm shift in artificial intelligence, moving beyond task-specific applications to powering autonomous agents capable of interacting with complex digital and physical environments. This new generation of "physical AI" holds immense promise for advancing robotics, industrial control, and laboratory automation by translating high-level human intent into executable code. However, the transition from virtual tasks to real-world action exposes a critical challenge: ensuring that the generated code is not merely syntactically valid, but functionally correct and safe. An error in a chatbot is an inconvenience; an error in controlling a costly and scarce scientific instrument can cause severe equipment damage and result in the loss of invaluable experimental time.

While the use of LLMs for code generation~\cite{jiang2024survey, chen2021evaluating} is growing, evaluation metrics are continually evolving in an effort to more accurately measure correctness.
When at least one ground-truth example is available, reference-based evaluation metrics, such as BLEU~\cite{papineni2002bleu} or variants CrystalBLEU \cite{eghbali2022crystalbleu}, often fail to represent functional correctness and can thus be misleading. 
CodeBLEU~\cite{ren2020codebleu} leverages the n-gram matching capability of BLEU while also incorporating code syntax through abstract syntax trees (AST) and capturing code semantics via data-flow analysis. CodeBERTScore is a metric for evaluating code generation by leveraging pretrained models to independently encode both the generated and reference code, along with optional natural language context~\cite{zhou2023codebertscore}.
However, these reference-based approaches, whether with tokens or embeddings, can still be misleading when the goal is to assess true functional correctness~\cite{chen2021evaluating, naik2024limitations}. 
Without the need for unit tests and reference code, metrics that use LLMs as judges can provide improved correlations with functional correctness and human preferences~\cite{zhuo2023ice, tong2024codejudge}.
%
Functional correctness is assessed by executing generated code and checking if its outputs match expectations, ensuring alignment with human judgment and real-world relevance.
This is most commonly done through unit tests, which compare the program’s output against predefined results for given inputs. Such approaches are standard in coding competitions, where human-written code is validated using a combination of public and hidden test cases.
Pass@$k$ gives the expected fraction of problems for which at least one of $k$ independently sampled completions passes all unit tests~\cite{kulal2019spoc}. 
Even execution-based benchmarks like APPS~\cite{hendrycksapps2021, vandervleuten2023drbootbootstrappingprogram} , which verify generated code through hidden test cases, are limited to stateless, algorithmic tasks and cannot capture more complex dynamics or continuous interaction with a changing environment.
SWE-Bench~\cite{jimenez2023swe} evaluates the LLM ability to reason and modify the codebase for realistic GitHub issues, while AgentBench~\cite{liu2023agentbench} benchmarks LLMs on their reasoning and decision-making abilities as agents in multi-turn, open-ended real-world environments.
These evaluations reflect a growing shift toward execution- and interaction-based methods for assessing LLMs.

Providing a sufficiently realistic execution environment for code evaluation requires the development and use of practical mechanisms such as mocking frameworks or digital twins. 
Digital twins have become increasingly central to scientific and industrial applications, serving as high-fidelity virtual representation of physical systems used for simulation, monitoring, and optimization \cite{rasheed2020digital, stadtmann2024physics, van2022executable}. 
Developing digital twins is complex and time-intensive, requiring expert-crafted models while facing challenges such as data sparsity and trade-offs between accuracy, efficiency, and interpretability in physics-based and data-driven approaches.
Evaluating digital twins remains challenging owing to their dynamic behavior and the multi-faceted nature of assessing their fidelity and performance, prompting studies to investigate various strategies.
\citet{ershenko2025quantitative} conducted a comparative study of error-based metrics on a digital twin of a railway braking system, showing that the appropriate metric is task specific: aggregate percentage-based measures such as mean absolute percentage error (MAPE) are well suited for judging overall fidelity, while instantaneous error measures like absolute percentage error (APE) are critical for triggering timely interventions. Complementing this metric-based perspective, 
\citet{munoz2022using} introduced a trace-alignment approach to assess behavioral similarity between digital and physical twins using offline execution traces. \citet{munozMeasuringDigitalTwins2024} also demonstrated runtime fidelity monitoring through continuous trace comparison. Trace-alignment methods have also been widely explored in process mining, where techniques derived from sequence-alignment are used to measure deviations between observed and expected behavior~\cite{weerdt2012processalignment, yang2017pima}. 
Together, these studies underscore the importance of dynamic, behavior-based evaluation for digital twins. Our work extends such behavioral evaluation to the domain of code generation, assessing the functional correctness of LLM-generated control logic through its interaction with a simulated environment.

While digital twins provide realistic, physics-based environments for evaluation, synergistically LLMs can also enable efficient control and design of digital twins, progressing toward agentic orchestration of experiments. 
Advances in LLMs enable automated code generation for simulations, control, and data pipelines, reducing manual effort and enhancing the adaptability and scalability of digital twins~\cite{yang2025leveraging, zhang2024large, wang2025llm, lin2024ddd}. 
LLM-driven digital twins or experimentation show strong potential for complex system reasoning, but deploying them safely in critical domains requires strict safety measures, expert oversight, and tailored evaluation methods.
Recently, \citet{hellert2025agentic, hellert2025alpha} presented an agentic system for a multi-stage physics experiment at a synchrotron light source storage ring to significantly reduce the preparation time.
\citet{chen2025agentic} demonstrated an agentic AI workflow capable of conducting experiments, including planning, execution, analysis, and iterative refinement to achieve a scientific objective in simulation (with calculated diffraction intensity) and also experimentally. The work further emphasizes the need for simulated environments to safely develop and deploy agentic systems. 
Other works have also explored the incorporation of language models for physical experiments~\cite{yao2025operationalizing, mathur2024vision, sulc2024towards, prince2023opportunities, tsai2023vision}, highlighting the increasing interest in agentic experimental workflows in science.

We have previously demonstrated a modular AI assistant VISION~\cite{mathur2024vision} that can voice-control synchrotron~\cite{willmott2019introduction} beamlines by generating Python code from natural language queries. 
Although we explored quantitative evaluation based on existing code similarity metrics, including exact string matching, Levenshtein distance, and CodeBLEU \cite{ren2020codebleu}, these syntactic metrics are limited in capturing functional correctness, as they often penalize correct code with differing structure or style.
For example, a \texttt{for} loop may be penalized when written in place of an equivalent \texttt{while} loop; semantic bugs can go undetected when the structure is correct but the parameters are wrong, such as moving a motor to \texttt{+10} rather than \texttt{-10}.
This gap between syntactic similarity and functional correctness is the primary barrier to the safe and reliable deployment of LLM agents in high-stakes physical systems.
To address these limitations, here we present EnvTrace\footnote{\url{https://github.com/CFN-softbio/EnvTrace}}, a semantic code evaluation framework that focuses on assessing functional correctness and runtime performance based on a program’s environmental effects, rather than its syntax. Leveraging a control-logic digital twin, the framework evaluates code behavior, with outcomes cross-validated against human judgment to ensure alignment.
The digital twin is deployed within a secure, sandboxed environment, providing a controlled and risk-free setting.

We detail the mock simulation environment and quantitative multi-faceted evaluation in Section~\ref{sec:methods}, evaluation of coding outcomes for multiple open- and closed-source LLMs in Section~\ref{sec:results}, discussion and summary in Sections~\ref{sec:discussion} and ~\ref{sec:conclusion}, respectively.

\begin{figure}[htbp]
\centering
\begin{minipage}{0.8\textwidth} 
\centering

\begin{tikzpicture}[
    node distance = 3mm and 35mm,
    every node/.style = {font=\sffamily\footnotesize},
    base/.style = {
      text width = 30mm,
      align = left,
      rounded corners = 2pt,
      inner sep = 2pt,
      minimum height = 1.8em,
      drop shadow = {shadow xshift=0.3pt,shadow yshift=-0.3pt,opacity=0.3}
    },
    gtpr/.style  = {base, draw=emerald!70!black, fill=emerald!8, text=emerald!80!black, line width=0.8pt},
    mismatch/.style = {base, draw=crimson!80!black, fill=crimson!8, text=crimson!80!black, line width=0.8pt},
    extra/.style = {base, draw=slate!60, fill=slate!3, text=slate!70, dashed, line width=0.6pt},
    blank/.style = {minimum width=30mm, minimum height=1.8em, inner sep=0pt},
    matchline/.style = {emerald!70!black, line width=1.2pt},
    misline/.style   = {crimson!70!black, line width=1.2pt, dashed},
]

\matrix (legend) [
  matrix of nodes,
  nodes={anchor=center,font=\sffamily\scriptsize,text depth=0pt},
  column sep=3mm,
  anchor=west
] {
  \node[gtpr, minimum width=5mm, minimum height=1em, text width=5mm] {}; & 
  \node {Matched event}; & 
  \node[mismatch, minimum width=5mm, minimum height=1em, text width=5mm] {}; & 
  \node {Value mismatch}; & 
  \node[extra, minimum width=5mm, minimum height=1em, text width=5mm] {}; & 
  \node {Extra LLM code event}; \\
};
\begin{scope}[on background layer]
  \node[
    fit=(legend),
    draw=slate!30,
    fill=white,
    rounded corners=2pt,
    inner sep=3pt,
    drop shadow = {shadow xshift=0.2pt,shadow yshift=-0.2pt,opacity=0.2}
  ] {};
\end{scope}

\node[font=\sffamily\bfseries\small, 
      fill=slate!10, 
      rounded corners=2pt, 
      inner sep=3pt,
      minimum width=30mm,
      below=8mm of legend.south west,
      anchor=north west] (hdrGT) {Ground Truth};
\node[font=\sffamily\bfseries\small, 
      fill=slate!10, 
      rounded corners=2pt, 
      inner sep=3pt,
      minimum width=30mm,
      right=of hdrGT] (hdrPR) {LLM Code};

\node[gtpr, below=4mm of hdrGT] (g1) {\texttt{AcquireTime = 1.0}};
\node[gtpr, right=of g1]        (p1) {\texttt{AcquireTime = 1.0}};
\node[gtpr, below=of g1] (g2) {\texttt{Acquire = 1}};
\node[gtpr, right=of g2] (p2) {\texttt{Acquire = 1}};
\node[gtpr, below=of g2] (g3) {\texttt{Acquire = 0}};
\node[gtpr, right=of g3] (p3) {\texttt{Acquire = 0}};
\node[gtpr, below=of g3] (g4) {\texttt{X\_Mtr = -0.2}};
\node[gtpr, right=of g4] (p4) {\texttt{X\_Mtr = -0.2}};
\node[gtpr, below=of g4] (g5) {\texttt{Z\_Mtr = -0.1}};
\node[gtpr, right=of g5] (p5) {\texttt{Z\_Mtr = -0.1}};
\node[blank, below=of g5] (gBlank1) {};
\node[extra, right=of gBlank1] (pIns1) {\texttt{Acquire = 1}};
\node[blank, below=of gBlank1] (gBlank2) {};
\node[extra, right=of gBlank2] (pIns2) {\texttt{Acquire = 0}};
\node[gtpr, below=of gBlank2] (g6) {\texttt{Z\_Mtr = 0.0}};
\node[mismatch, right=of g6]   (p6) {\texttt{Z\_Mtr = 0.2}};
\node[gtpr, below=of g6] (g7) {\texttt{X\_Mtr = -0.2}};
\node[gtpr, right=of g7] (p7) {\texttt{X\_Mtr = -0.2}};

\foreach \i in {1,...,5,7} {
  \draw[matchline] (g\i.east) -- (p\i.west);
}
\draw[misline] (g6.east) -- (p6.west);

\end{tikzpicture}

\vspace{5mm}

\end{minipage}

\caption{Illustration of the EnvTrace alignment process and score calculation. The ground-truth and LLM code (predicted code) execution traces are aligned based on their Process Variable (PV) changes. Matched events are shown by connected solid lines, value mismatches by dashed lines, and extra predicted events are shown in dashed box. This alignment is then used for computing scores to evaluate the LLMs.
}
\label{fig:trace-alignment}
\end{figure}
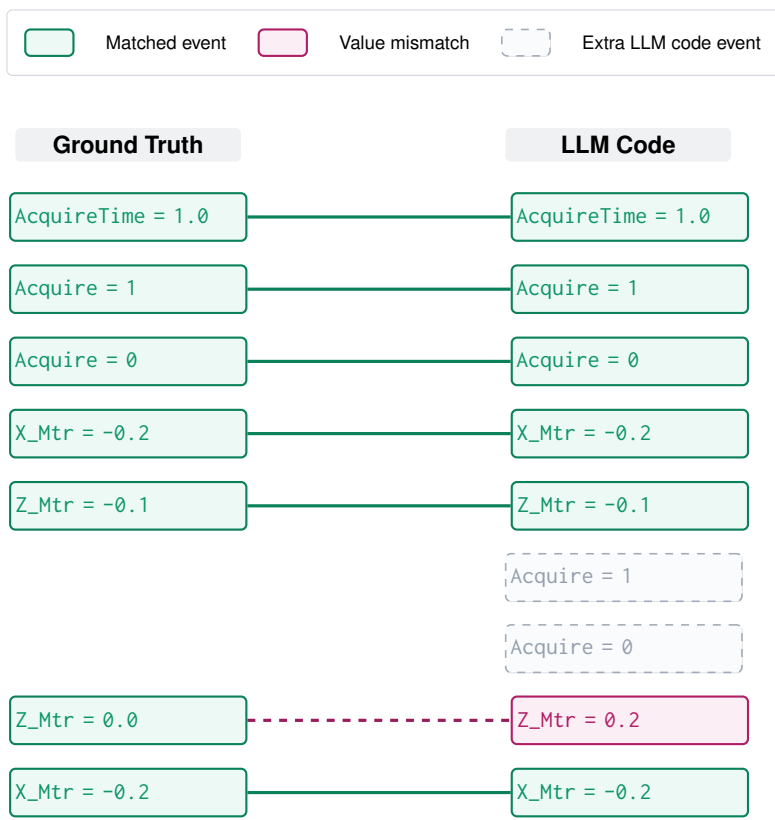


\section{Methodology}
\label{sec:methods}

EnvTrace overcomes the limitations of syntactic code evaluation by assessing the semantic and functional correctness of LLM-generated code in a simulated physical environment. 
It involves the execution of both the ground-truth (reference) code and the LLM-generated code within the mock environment or simulator. An execution trace, which is a time-ordered log of all state changes to the environment's parameters, is captured for each code. By systematically aligning and comparing these traces, as shown in Fig.~\ref{fig:trace-alignment}, we can determine if the two pieces of code produced the same functional outcome, executed in the correct sequence, and with correct timing.
While \citet{chen2025agentic} utilizes calculated diffraction intensity for instrument feedback, our work emphasizes tracking simulated physical instruments, offering a valuable complementary perspective.
Here we demonstrate and validate EnvTrace at synchrotron beamlines~\cite{willmott2019introduction}, where extremely intense X-ray enables studies in e.g. microelectronics~\cite{aidukas2024high, holler2019three, holler2017high}, energy and material science~\cite{sidhik2024two, xu2024unraveling}, and biomedical science~\cite{xiao2025application, shahmoradian2017three}.
The beamline environment is a quintessential cyber-physical system: it is highly stateful, where actions have persistent effects; it is real-time, where the timing and pacing of operations are critical; and its control involves a complex interplay of discrete actions (e.g., motor movements, temperature targets, shutter triggers) and continuous processes (e.g., temperature ramps). The high-stakes nature of synchrotron experimentation, due to the limited experiment time and expensive instrumentation, makes it a powerful and representative domain for developing evaluation methods that can be generalized to other physical AI systems.

The simulation-based evaluation paradigm offers several key advantages over static code analysis. First and foremost, it enables a true functional equivalence comparison. Instead of a binary pass/fail score from an exact string match or a similarity score from CodeBLEU, this approach provides a granular, multi-dimensional evaluation that compares code on an action-by-action basis. 
Second, the framework provides scientists and developers with a valuable resource for debugging code, exploring alternative implementations, and validating ground-truth examples, thereby strengthening the testbed.
Third, the simulator enables "pre-flight check" of experimental plans, allowing researchers to detect errors in complex sequences before using limited and costly experiment time, e.g. beamtime at X-ray or neutron facilities. This highlights the value of robust digital twins for developing safe and efficient agent-based systems in scientific research.

\subsection{Simulation Environment}
\label{sec:simulation_env}

To enable execution-based evaluation without risking physical hardware or consuming scarce beamtime, we developed a control-logic beamline simulator, practical for realistic instrument-code evaluation.
The simulator provides the actual beamline environment through scripts defined in the Python-based Bluesky data acquisition framework~\cite{bluesky}. These scripts not only specify the basic Bluesky configurations, but also capture beamline-specific devices and measurement protocols, enabling faithful reproduction of experimental workflows in a simulated setting. 
As Bluesky continues to be adopted across synchrotron facilities, such a simulator may gain in broad applicability and suitability for authentic assessment of instrument-code.

\begin{figure}[hb]
  \centering
  \includegraphics[width=\textwidth]{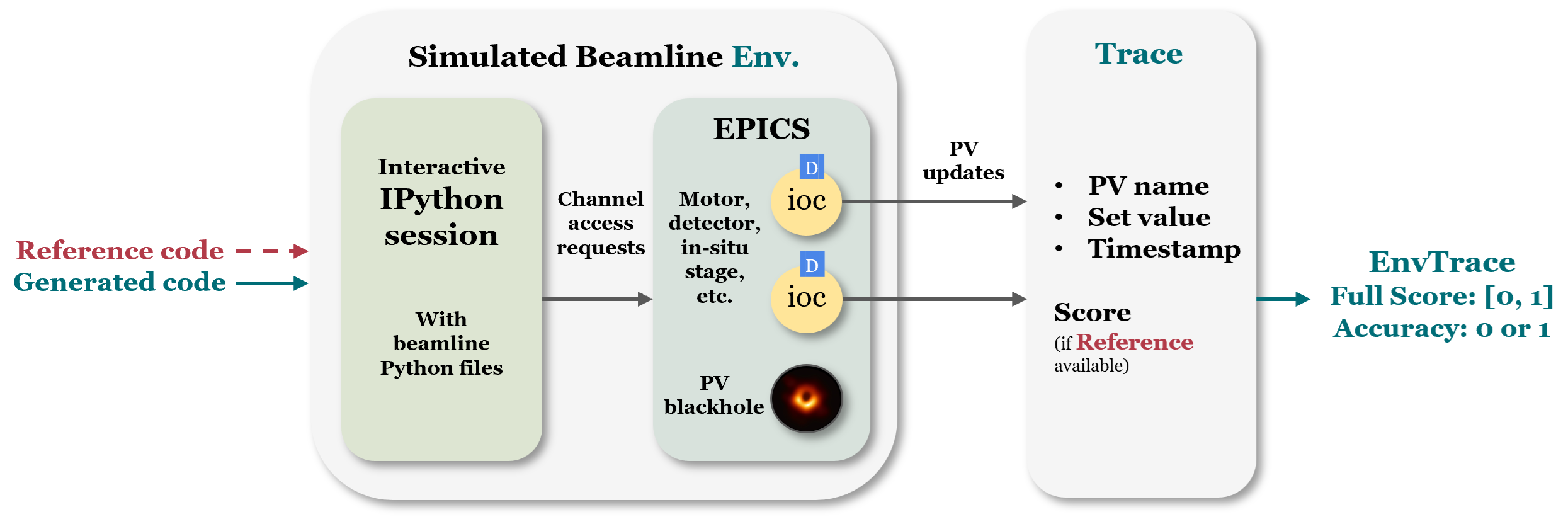}
  \caption{An overview of the EnvTrace architecture. LLM-generated code and ground-truth code are executed within an interactive IPython session. 
  This session communicates with a simulated beamline environment, which runs EPICS IOCs in Docker containers. An EPICS monitor captures all state changes and generates an execution trace. 
  The ground-truth and predicted code traces are then aligned to compute a multi-faceted EnvTrace full score based on state, temporal, and behavioral equivalence\protect\footnotemark.}
  \label{fig:methods_overview}
\end{figure}

An overview of the EnvTrace implementation for beamline code evaluation is provided in Fig.~\ref{fig:methods_overview}. Two sets of code, ground-truth/reference and human- or LLM-generated, are executed at the simulated beamline. During execution, the monitoring system logs the states of beamline components to produce a multi-faceted score that quantifies the functional similarity between the two.
This control-logic digital twin replicates the control infrastructure of the beamline by modeling relevant components with configurable physical accuracy, including motors, thermal stage, and detectors, while ignoring non-essential commands by mocking them through a black hole IOC (Input-Output Controller).
As shown in Fig.~\ref{fig:methods_overview}, the simulator is composed of two subsystems: the interactive control session and the simulated Experimental Physics and Industrial Control System (EPICS)~\cite{dalesio2020epics} services. 
Our simulator replicates this using a set of Docker containers and Python scripts, each running an EPICS IOC.
The fidelity of our control-logic digital twin is specifically tailored for validating control logic rather than performing a full physics-based simulation. The IOCs achieve this by providing two key levels of abstraction:
(1) Interface Replication: They expose the exact same EPICS Process Variables (PVs) as the physical hardware.
(2) State Tracking: They maintain a persistent internal state for critical components. For example, the simulated motor IOCs track and update their position values after a move command, ensuring that subsequent commands operate on the correct state.
The simulation does not, however, reproduce more complex realistic behaviors such as generating synthetic detector images or modeling the precise thermodynamics of a temperature ramp. This level of functional, state-aware abstraction is sufficient for verifying the correctness of command sequences, parameters, and timing.
The mock framework executes code within a sandboxed IPython session, communicating with simulated devices via the standard EPICS Channel Access network protocol.
This ensures that, from the control software’s perspective, interactions with the simulated environment are functionally indistinguishable from those with real beamline hardware, allowing the same code to run in both simulated and live instrument control environments.
The detailed EnvTrace architecture is provided in Fig.~\ref{si:overview}, with the implementation for the beamline control-logic simulator available at: \url{https://github.com/CFN-softbio/VISION}.

In this work, we demonstrate EnvTrace within the beamline environment at the National Synchrotron Light Source II (NSLS-II) 11-BM Complex Materials Scattering (CMS) beamline.
The simulator can run the same Python code used in actual beamline operations without modification, making it easily adaptable for deployment at other beamlines. This architecture generates realistic sequences of state changes that form the basis of our evaluation.

\footnotetext{The black hole image is courtesy of the Event Horizon Telescope Collaboration \cite{theeventhorizontelescopecollaborationFirstM87Event2019} (CC BY 4.0).}

\subsection{Multi-Faceted Evaluation}
\label{sec:envtrace_framework}

Built upon this simulation environment, EnvTrace provides a methodology for assessing the semantic equivalence of code snippets by comparing their execution traces.
An execution trace is a time-ordered sequence of state changes observed in the environment. Each entry in the trace is a tuple containing the Process Variable’s (PV) name, its new value, and a high-precision timestamp:
$(\text{pv\_name}, \text{value}, \text{timestamp})$.
The process begins when a code snippet, either ground-truth or LLM-generated, is passed to the evaluation framework, as illustrated in Fig.~\ref{fig:methods_overview}.
The code is injected into the simulated IPython session, where its execution modifies the state of the simulated IOCs. An EPICS monitor, implemented via the \texttt{camonitor} utility, captures all PV updates and records them to construct the execution trace.
Specific function calls can be selected or omitted for tracking.

Comparing two execution traces requires a more nuanced approach than a simple one-to-one check. EnvTrace computes a holistic \texttt{full\_score}, given in Eq.(\ref{eq:full_score}), by analyzing three distinct aspects of equivalence: 
\textbf{(1) State Sequence Alignment:}  
the first and most critical aspect is whether the correct sequence of operations occurred. We compare the ground-truth and the generated code traces using a sequence alignment algorithm from \texttt{difflib.SequenceMatcher}. 
This algorithm identifies matching subsequences and highlights insertions, deletions, and substitutions. 
Using this alignment, we compute a \texttt{pv\_match\_rate}, which represents the fraction of operations that are both of the expected PV type and contain the correct values.
\textbf{(2) Temporal Dynamics Analysis:}
for operations executed in the correct sequence, it is equally important that they occur with the proper timing and pace. Using the set of correctly matched PV changes from the alignment step, corresponding timestamps were extracted. A linear regression is performed between timestamps of the ground-truth and predicted code. A high coefficient of determination ($R^2$) indicates consistent pacing, while a slope close to 1.0 indicates a correct overall duration. These factors, along with the Mean Absolute Percentage Error (MAPE) of the time intervals, are combined into a \texttt{timing\_score}.
\textbf{(3) Continuous Process Fidelity:}
Some beamline operations, like temperature ramping, are continuous processes where the evolution of a PV state over time is critical, not just its final state. To evaluate these, EnvTrace employs "process fidelity metrics" designed to compare the overall trajectory of these values. For instance, when evaluating a temperature ramp, we calculate the Mean Absolute Error (MAE) across the entire profile, as well as the absolute difference in the final temperature. These are combined into an exponentially decaying \texttt{temp\_score}, which rewards close tracking of the target profile.
The specific thresholds and detailed logic for each component are provided in Section~\ref{si:continuous_score_formulas}.

The motor PVs and temperature are monitored in this work, however the metric can be adapted to accommodate other instrument status, e.g. in-situ sample humidity or pressure.
Here three individual scores are combined into a single, weighted simulation \texttt{full\_score} that provides a comprehensive measure of functional equivalence,
\begin{equation}
\label{eq:full_score}
\texttt{full\_score} = 
\begin{cases} 
    0.6 \cdot \texttt{pv\_match\_rate} + 0.2 \cdot \texttt{timing\_score} + 0.2 \cdot \texttt{temp\_score} & \text{if temperature} \\
    0.8 \cdot \texttt{pv\_match\_rate} + 0.2 \cdot \texttt{timing\_score} & \text{if no temperature}
\end{cases}
\end{equation}
    
This weighting is chosen to reflect a hierarchy of importance for physical systems. The state sequence, captured by the \texttt{pv\_match\_rate}, receives the highest weight as it represents the \textit{primary} functional correctness, i.e., whether the code produce the correct operations with the correct targeted values. A failure in this aspect constitutes a fundamental error.
In contrast, the \texttt{timing\_score} and \texttt{temp\_score} represent \textit{secondary} performance characteristics. This is because multiple, equally valid interpretation of intent and coding strategies can exist to achieve the same high-level experimental goal, each producing a different low-level execution trace. For instance, a task to ``measure every 20$^{\circ}$C during a ramp to 200$^{\circ}$C'' could be implemented in at least two functionally correct ways:
(a) By setting a single final temperature target of 200$^{\circ}$C and programming a loop that waits for the temperature to cross multiples of 20$^{\circ}$C before triggering a measurement.
(b) By programming a loop that explicitly sets a series of discrete 20$^{\circ}$C heating steps, e.g., set to 20$^{\circ}$C, wait, measure; set to 40$^{\circ}$C, wait, measure; and so on.
While both strategies achieve the desired instrument and measurement outcomes, different PV traces are generated for the temperature setpoint and result in varying temporal profiles. 
Consequently, the scoring metric can be defined to prioritize the successful execution of the core commands, such as the sequence of detector triggers that takes measurements, 
which are evaluated by the highly-weighted \texttt{pv\_match\_rate}. The specific implementation path, reflected in the temperature profile and timing, is evaluated by the lower-weighted \texttt{temp\_score} and \texttt{timing\_score}. This approach ensures that implementations which should be considered functionally correct yet logically or stylistically different still achieve a high overall score, recognizing them as valid solutions.
This EnvTrace full score, ranging from 0.0 to 1.0, forms the basis of our semantic evaluation. When there are multiple reference answers, the one that gives the highest full score is used to assess the model performance.

As a strict and conservative metric, we also employed a measure referred to as `EnvTrace accuracy', which is a more stringent, composite criterion with only binary results (success or failure).
This also provides a comparison with the binary string-based exact match approach as used in \citet{mathur2024vision}. 
Instead of applying a threshold to each score component in Eq.~(\ref{eq:full_score}) as in the full score, the generated code snippet is considered a functional match with a score of 1 only if it passes a series of independent strict checks across all relevant aspects of its execution trace, otherwise it is considered a mismatch with a score of 0.
EnvTrace accuracy serves as a simulator informed lower-bound due to its strictness.
This metric for evaluating functional match requires an exact match in the sequence of state changes, strict adherence to temporal dynamics, and, when relevant, high accuracy in tracking continuous processes like temperature ramps. 
Any deviation from these criteria results in the code being categorized as a mismatch. 
Although some may view this metric as overly conservative, it provides an additional means of evaluation, particularly for highly sensitive systems that demand an extremely high degree of functional and semantic correctness. Details for defining accuracy are given in Section~\ref{si:accuracy_details}.

\section{Results}
\label{sec:results}

Our investigation and evaluation are structured around two central questions: (1) Does the proposed semantic, execution-based approach yield a more reliable measure of code correctness than traditional syntactic metrics? and (2) How do state-of-the-art LLMs perform under this more rigorous benchmark for instrument control? 
The benchmark dataset for evaluating LLM performance includes
116 simple cases comprising single- or few-line commands and
20 complex-flow control examples involving \texttt{for} or \texttt{while} loops.
The ground-truth (GT) solutions were generated through a collaborative process involving two human scientists and \texttt{Claude-3.7-Sonnet}. Each GT entry was executed within our simulator to verify its functional correctness and resolve bugs. Due to the inherent ambiguity in some natural language prompts, most examples have multiple valid GT implementations, therefore a functional or exact string match with any of them is considered correct. 
Ground-truths here should be viewed as reference implementations, and due to inherent ambiguity in intent, a less-than-perfect full score may reflect an interpretation of intent mismatch rather than an actual error. 
The resulting scores should be interpreted in this context, while also considering that fewer than five GTs are present per example. 
Moreover, to better interpret the full score of LLMs, two human beamline scientists also provided their code implementations for the complex-flow prompts. 
Human submissions were manually debugged using the simulator to reflect typical beamline workflows, and the corrected versions were evaluated with EnvTrace.

Our study benchmarked the code quality and performance across closed- and open-source models, with model details provided in Section~\ref{si:models}.
For all LLM evaluations, each prompt was run 3 times and the results averaged to strengthen statistical significance.
To improve efficiency when evaluating across datasets and models, the framework employs caching to avoid redundant executions. Once a piece of code has been run in the simulator, subsequent runs retrieve the stored PV changes directly, rather than re-executing the simulation.
The variability in evaluation results may arise from the stochastic nature of LLM code generation and from fluctuations in simulator execution timing. With the use of caching, the effect of simulator variability will be less visible in the results.
The scoring outcomes are thus also discussed qualitatively along with the quantitative results.

\subsection{Model Performance on Simple-Flow Tasks}
\label{sec:simple_flow_results}

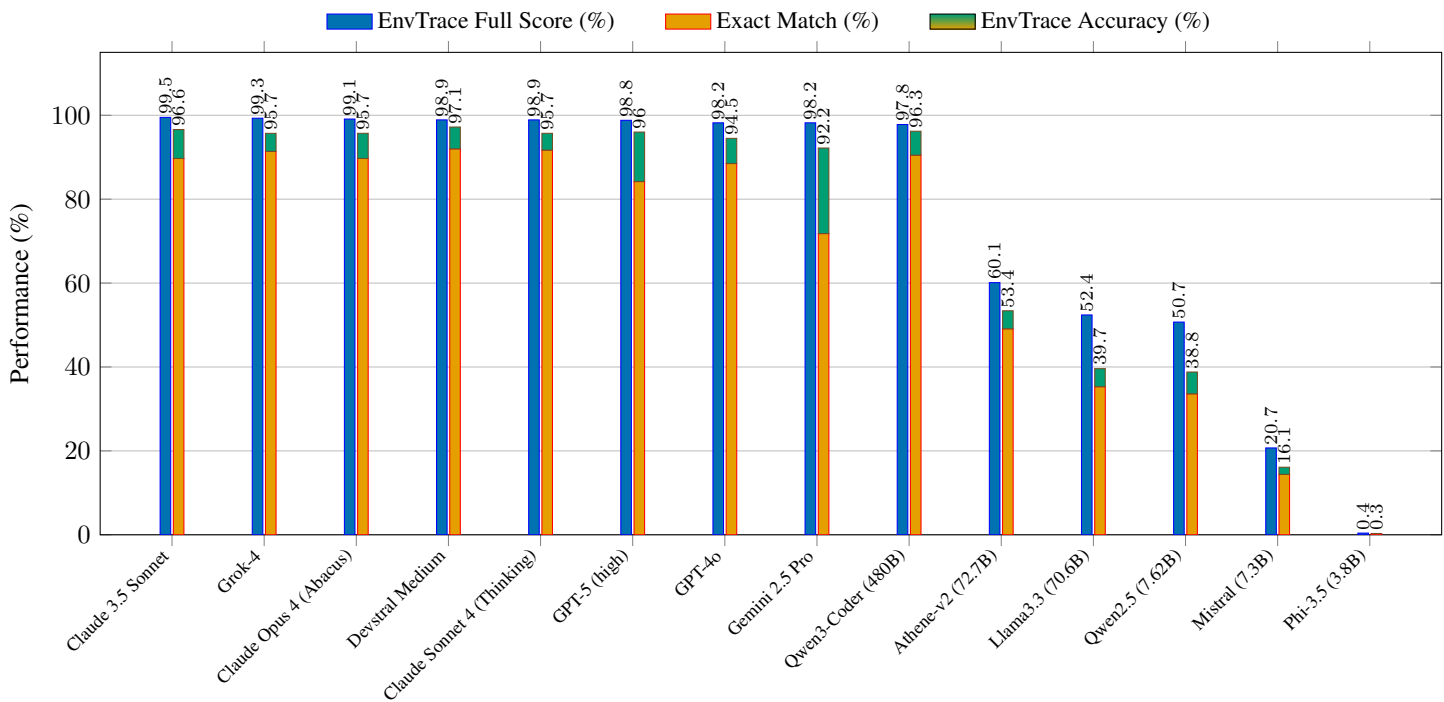
\begin{figure}[ht]
\makebox[\textwidth][c]{%
\begin{minipage}{1.12\textwidth}
\centering
\caption{Selected model performance on simple-flow tasks in Table~\ref{tab:simple_results}. Left blue bars show the EnvTrace full score, and right stacked bars give the string-based exact match (bottom, orange) and improvement (top, green) with EnvTrace accuracy.}
\label{fig:simple_tasks_performance}

\begin{tikzpicture}
\begin{axis}[
    width=1.05\textwidth, height=8cm,
    ybar,
    bar width=4pt,
    enlarge x limits=0.06,
   symbolic x coords={ {Claude 3.5 Sonnet},{Grok-4},{Claude Opus 4 (Abacus)},{Devstral Medium},{Claude Sonnet 4 (Thinking)},{GPT-5 (high)},{GPT-4o},{Gemini 2.5 Pro},{Qwen3-Coder (480B)},{Athene-v2 (72.7B)},{Llama3.3 (70.6B)},{Qwen2.5 (7.62B)},{Mistral (7.3B)},{Phi-3.5 (3.8B)} },
    xtick=data,
    xticklabel style={rotate=45,anchor=east,font=\scriptsize},
    ylabel={Performance (\%)},
    ymin=0, ymax=115,
    ymajorgrids=true,
    xmajorgrids=false,
    legend style={
        at={(0.5,1.02)}, anchor=south,
        legend columns=-1, /tikz/every even column/.append style={column sep=0.6cm},
        draw=none, fill=none, font=\small
    },
]

\addplot+[area legend, fill=oiBlue, bar shift=-2.5pt,
          nodes near coords,
          every node near coord/.append style={font=\scriptsize,rotate=90,anchor=west,inner sep=1pt,text=black},
          point meta=y]
    table[x=label,y=fullscore]{\datatable};
\addlegendentry{EnvTrace Full Score (\%)}

\addplot+[area legend, ybar, stack plots=y, fill=oiOrange, bar shift=+2.5pt]
    table[x=label,y=exact]{\datatable};
\addlegendentry{Exact Match (\%)}

\addplot+[ybar, stack plots=y, fill=oiBluishGreen, bar shift=+2.5pt, forget plot]
  table[x=label,y=improvement]{\datatable};

\addlegendimage{area legend, ybar, draw=none, shade,
  bottom color=oiOrange, top color=oiBluishGreen}
\addlegendentry{EnvTrace Accuracy (\%)}

\addplot+[draw=none, fill=none, bar shift=+2.5pt,
          nodes near coords, point meta=explicit,
          every node near coord/.append style={
            font=\scriptsize, rotate=90, anchor=west, inner sep=1pt, text=black
          }]
    table[x=label,y=total,meta expr=\thisrow{total}]{\datatable};

\end{axis}
\end{tikzpicture}
\end{minipage}%
}
\end{figure}

EnvTrace was used for LLM evaluation on simple-flow examples, which typically involve single-line commands for actions, e.g. moving a motor or taking a measurement.
The performance of 31 LLMs (including reasoning and vendor variants) on the 116 simple flow tasks is detailed in Table~\ref{tab:simple_results}, with selected model performance plotted in Fig.~\ref{fig:simple_tasks_performance}.
In each model family, the model performance is sorted by the EnvTrace full score, from highest to lowest. There is a notable difference between string-based `exact match' and `EnvTrace accuracy'.
This difference arises because LLMs often generate code that is functionally equivalent but syntactically distinct from the ground-truth examples. 
Common variations include using different variable names, adding comments, explicitly naming default parameters in functions, or  using different functions to accomplish the same behavior. 
While these variations cause syntactic mismatch, EnvTrace can correctly identify them as functionally identical by verifying that they produced the same sequence of state changes in the simulator. 
As shown in Fig.~\ref{fig:simple_tasks_performance}, each LLM's performance is shown with the blue bar (on the left) representing the EnvTrace full score, while the orange and green stacked bar (on the right) reflects exact match accuracy and the added correctness captured by the simulator, respectively. The green segment reflects improvement in functional correctness missed by syntactic metrics.

An example of a simple-flow task, with corresponding LLM results and evolution, is given in Box~\ref{lst:simple}. 
The natural language query ``Move the organic thin film x by 5.8 and align the thin film'' corresponds to moving the sample in x direction with \texttt{sam.xr(5.8)} and perform the sample alignment routine with \texttt{sam.align()}. 
If the generated code performs the correct actions but additionally specifies a sample name as by \texttt{Claude-Sonnet-4-(Thinking)}, string-based exact match would fail, while EnvTrace would return a perfect PV match, achieved EnvTrace accuracy, and a full score ideally 100\%. 
In the case where the generated code is partially correct, as in the case of \texttt{Qwen2.5-Coder} where the alignment function is written incorrectly, EnvTrace can indicate a partial success by giving a score of 60\%.
EnvTrace accuracy and full score, both stemmed from the same trace-based semantic analysis, provide improved and meaningful indicators of LLM coding performance on simple-flow tasks.
By performing a deeper trace analysis, the full score serves as a stronger proxy for performance.

\begin{lstlisting}[float, language=Python, caption={Example of a simple-flow prompt and LLM code, showing that while closed-source model \texttt{Claude-Sonnet-4 (Thinking)} provided an accurate solution, the open-source model \texttt{Qwen2.5-coder} hallucinated one of the commands.}, label={lst:simple}]
NL INPUT: Move the organic thin film x by 5.8 and align the thin film
------------------------------------------------------
#GROUND-TRUTH SNIPPET: 
sam.xr(5.8); 
sam.align()
======================================================
#Claude Sonnet 4 (Thinking) PREDICTED SNIPPET:
sam = Sample('organic_thin_film')
sam.xr(5.8); 
sam.align()
#-----------------------------------------------------
#PV match rate: 100.00%
#Timing match: True (score: 0.999)
#EnvTrace Accuracy: True (Full score: 1.000)
======================================================
#Qwen2.5-coder PREDICTED SNIPPET:
sam.xr(5.8); 
sam.aligned()
#-----------------------------------------------------
#PV match rate: 50.00%
#Timing match: True (score: 1.000)
#EnvTrace Accuracy: False (Full score: 0.600)
\end{lstlisting}

Table~\ref{tab:simple_results} presents the performance of both closed- and open-source models, showing the EnvTrace full score and accuracy, alongside string-based exact match, normalized Levenshtein distance, and inference time for comparison.
Closed-source models from vendors such as Anthropic, xAI, and OpenAI form the top tier of performance, with all achieving greater than 90\% on EnvTrace accuracy and more than 95\% on full score. 
The less than perfect scores are generally due to, for example, the ground truth assuming that the thermal stage is already powered on when setting the temperature, but some LLMs turn it on again (redundant but harmless), resulting in an extra PV change and a slight mismatch. This illustrates the idea that an imperfect score can imply  `mismatch' rather than  `error'.
For some models, stochasticity can make a noticeable difference. 
For example, in a simple command to move the sample stage, \texttt{GPT-5-(minimal)} occasionally ‘over-thought’ the task and, instead of simply moving the motor, it asked for clarification on the exposure time and failed to generate any code. This results in a lower performance score compared to other models in the OpenAI family.
Meanwhile, \texttt{GPT-5-(high)} utilized \texttt{try-except} for robustness, although unnecessary in this case, but indicative of useful deep thinking for more complex scenarios.
While open-source model \texttt{Qwen3-Coder} achieved performance comparable to closed-source ones, other open-source models performed significantly worse, with most achieving under 60\% in both accuracy and full score. 
\texttt{Athene-v2} and its agent-tuned variant, achieve full score around 60\% (with accuracies around 53\%). Other popular models like \texttt{Llama3.3} and the \texttt{Qwen} series fall into a 40-50\% full score range (with 30-40\% accuracy range). This highlights that generating a simple/small piece of code that is syntactically and semantically correct remains a challenging task that is not yet mastered by most open-source models. These models often make small but critical errors: they might understand the task but hallucinate the command or its format, as illustrated by the \texttt{Qwen2.5-coder} example in Box~\ref{lst:simple}. 
Smaller open-source models like \texttt{Qwen2.5} and \texttt{Mistral-(7.3B)} offer fast inference (<1s), although at a considerable cost to accuracy. Closed-source model \texttt{GPT-4o} provides both fast inference (0.8s) and high accuracy (98\%).

To assess the impact of prompt length, we analyzed LLM performance with shorter prompts in Section~\ref{si:baseline_performance}.
When system prompts include more details and cover more functionalities, we observe that the performance of smaller LLMs decline significantly. For example, \texttt{Qwen2.5-Coder} performance dropped from a full score of over 98\% with the 2000 word prompts ($\sim$3800 tokens with \texttt{GPT-4o} tokenizer), as shown in Table~\ref{tab:simple_results_old_prompt_new_data}, to 38\% with the use of the 5000 word prompt ($\sim$8500 tokens) in Table~\ref{tab:simple_results}. This suggests that open-source models perform reasonably well when only limited knowledge is included in the system prompt; as the demand for broader knowledge or functionality increases, closed-source models lead in both accuracy and inference speed.

\begin{table}[htbp]
  \centering
  \begin{threeparttable}
    \caption{Performance of LLMs on Simple-Flow Code Generation Tasks (N=116)}
    \label{tab:simple_results}
    \sisetup{separate-uncertainty=true, table-align-uncertainty=true}
    \begin{tabular}{
      l
      S[table-format=2.2(2.2)]
      S[table-format=2.2(2.2)]
      S[table-format=2.2(2.2)]
      S[table-format=2.2(2.2)]
      S[table-format=2.2(2.2)]
    }
      \toprule
      \textbf{Model} & 
      {\textbf{\begin{tabular}[c]{@{}c@{}}EnvTrace \\ Full Score (\%, $\uparrow$)\tnote{a}\end{tabular}}} & 

      {\textbf{\begin{tabular}[c]{@{}c@{}} EnvTrace \\ Accuracy (\%, $\uparrow$)\tnote{b}\end{tabular}}} &
      {\textbf{\begin{tabular}[c]{@{}c@{}} \textcolor{gray}{Exact Match} \\  \textcolor{gray}{(\%, $\uparrow$)\tnote{c}}\end{tabular}}} & 
      {\textbf{\begin{tabular}[d]{@{}c@{}}\textcolor{gray}{Norm. Lev.} \\ \textcolor{gray}{Dist.} \textcolor{gray}{(\%, $\downarrow$)\tnote{d}}\end{tabular}}} &
      {\textbf{\begin{tabular}[c]{@{}c@{}}Inference \\ Time (s, $\downarrow$)\end{tabular}}} \\
\midrule
\multicolumn{6}{l}{\textbf{Closed Source Models}} \\
\midrule
\multicolumn{6}{l}{\textit{Anthropic}} \\
  Claude 3.5 Sonnet & 99.5 \pm 0.0 & 96.6 \pm 0.0 & 89.7 \pm 0.0 & 2.4 \pm 0.0 & 1.6 \pm 0.0 \\
  Claude Sonnet 4 (Bedrock) & 99.2 \pm 0.0 & 97.4 \pm 0.0 & 94.0 \pm 0.0 & 1.6 \pm 0.0 & 28.8 \pm 0.0 \\
  Claude Sonnet 4 (Abacus) & 99.2 \pm 0.0 & 97.4 \pm 0.0 & 94.0 \pm 0.0 & 1.6 \pm 0.0 & 2.6 \pm 0.1 \\
  Claude Opus 4 (Abacus) & 99.1 \pm 0.0 & 95.7 \pm 0.0 & 89.7 \pm 0.0 & 2.3 \pm 0.0 & 2.9 \pm 0.1 \\
  Claude Sonnet 4.5 (Thinking)\tnote{t} & 99.1 \pm 0.0 & 95.7 \pm 0.9 & 94.0 \pm 2.3 & 2.2 \pm 1.1 & 7.1 \pm 0.8 \\
  Claude Sonnet 4 (Thinking)\tnote{t} & 98.9 \pm 0.8 & 95.7 \pm 1.7 & 91.7 \pm 1.8 & 1.9 \pm 0.7 & 5.0 \pm 0.2 \\
  Claude Opus 4 (Thinking)\tnote{t} & 98.4 \pm 1.0 & 95.4 \pm 1.0 & 88.2 \pm 1.3 & 2.6 \pm 0.5 & 5.7 \pm 0.1 \\
\midrule
\multicolumn{6}{l}{\textit{xAI}} \\
  Grok-4\tnote{t} & 99.3 \pm 0.0 & 95.7 \pm 0.0 & 91.4 \pm 1.5 & 2.1 \pm 0.1 & 16.0 \pm 0.8 \\
  Grok-4 Fast & 98.8 \pm 0.6 & 95.7 \pm 2.6 & 89.7 \pm 0.9 & 2.5 \pm 0.2 & 3.4 \pm 1.2 \\
  Grok Code Fast 1 & 97.5 \pm 0.9 & 91.4 \pm 2.6 & 81.6 \pm 3.5 & 5.4 \pm 1.1 & 2.9 \pm 0.1 \\
\midrule
\multicolumn{6}{l}{\textit{OpenAI / Microsoft}} \\
  o3 (high)\tnote{t} & 99.1 \pm 0.5 & 96.0 \pm 0.5 & 89.7 \pm 1.7 & 2.5 \pm 0.5 & 4.9 \pm 0.4 \\
  GPT-5 (high)\tnote{t} & 98.8 \pm 0.6 & 96.0 \pm 0.5 & 84.2 \pm 2.2 & 6.3 \pm 1.4 & 31.8 \pm 2.1 \\
  GPT-4o & 98.2 \pm 0.0 & 94.5 \pm 0.5 & 88.5 \pm 0.5 & 3.5 \pm 0.1 & 0.8 \pm 0.0 \\
  GPT-4o (Abacus) & 97.6 \pm 0.0 & 92.8 \pm 0.5 & 80.7 \pm 0.5 & 4.7 \pm 0.1 & 2.3 \pm 0.6 \\
  GPT-5 (minimal)\tnote{t} & 95.7 \pm 2.2 & 92.2 \pm 2.3 & 84.2 \pm 2.2 & 6.5 \pm 1.8 & 1.3 \pm 0.1 \\
\midrule
\multicolumn{6}{l}{\textit{Mistral}} \\
  Devstral Medium\tnote{t} & 98.9 \pm 0.4 & 97.1 \pm 0.5 & 92.0 \pm 0.5 & 2.9 \pm 0.5 & 2.0 \pm 1.4 \\
\midrule
\multicolumn{6}{l}{\textit{Google}} \\
  Gemini 2.5 Pro\tnote{t} & 98.2 \pm 0.4 & 92.2 \pm 1.5 & 71.8 \pm 0.5 & 7.0 \pm 0.1 & 5.1 \pm 0.0 \\
\midrule
\multicolumn{6}{l}{\textbf{Open Source Models}} \\
\midrule
  Qwen3-Coder (480B) & 97.8 \pm 0.7 & 96.3 \pm 1.0 & 90.5 \pm 0.9 & 3.2 \pm 0.7 & 2.0 \pm 0.5 \\
  Athene-v2-Agent (72.7B) & 62.9 \pm 0.0 & 52.6 \pm 0.0 & 46.6 \pm 0.0 & 23.0 \pm 0.0 & 5.3 \pm 0.0 \\
  Athene-v2 (72.7B) & 60.1 \pm 0.0 & 53.4 \pm 0.0 & 49.1 \pm 0.0 & 21.0 \pm 0.0 & 5.2 \pm 0.0 \\
  Qwen3-Coder (30.5B) & 54.7 \pm 0.5 & 48.0 \pm 0.5 & 39.7 \pm 0.0 & 23.7 \pm 0.0 & 1.9 \pm 0.0 \\
  Llama3.3 (70.6B) & 52.4 \pm 0.0 & 39.7 \pm 0.0 & 35.3 \pm 0.0 & 25.6 \pm 0.0 & 5.1 \pm 0.0 \\
  Qwen2.5 (7.62B) & 50.7 \pm 0.0 & 38.8 \pm 0.0 & 33.6 \pm 0.0 & 23.4 \pm 0.0 & 0.8 \pm 0.0 \\
  Qwen2 (7.62B) & 46.8 \pm 0.0 & 37.1 \pm 0.0 & 31.9 \pm 0.0 & 28.2 \pm 0.0 & 0.8 \pm 0.0 \\
  Qwen2.5-Coder (32.8B) & 37.6 \pm 0.0 & 31.9 \pm 0.0 & 28.4 \pm 0.0 & 32.0 \pm 0.0 & 2.7 \pm 0.0 \\
  GPT-oss (117B) & 33.2 \pm 0.0 & 27.6 \pm 0.0 & 25.9 \pm 0.0 & 59.8 \pm 0.0 & 240.8 \pm 0.3 \\
  Mistral-NeMo (12.2B) & 25.9 \pm 0.0 & 17.2 \pm 0.0 & 11.2 \pm 0.0 & 38.6 \pm 0.0 & 1.1 \pm 0.0 \\
  Mistral (7.3B) & 20.7 \pm 0.5 & 16.1 \pm 0.5 & 14.4 \pm 0.5 & 38.5 \pm 0.2 & 0.9 \pm 0.1 \\
  Phi-3.5 (3.8B, fp16) & 0.6 \pm 1.0 & 0.3 \pm 0.5 & 0.3 \pm 0.5 & 99.1 \pm 1.5 & 1.4 \pm 1.4 \\
  Phi-3.5 (3.8B) & 0.4 \pm 0.8 & 0.3 \pm 0.5 & 0.3 \pm 0.5 & 99.0 \pm 1.8 & 1.0 \pm 0.6 \\
  Qwen3 (32B) & 0.0 \pm 0.0 & 0.0 \pm 0.0 & 0.0 \pm 0.0 & 96.4 \pm 0.0 & 6.6 \pm 0.0 \\
\bottomrule
    \end{tabular}
    \begin{tablenotes}
      \small
      \item[a] \textbf{EnvTrace Full Score (\%):} The continuous score from 0 to 100, reflecting the weighted average of state and temporal fidelity.
      \item[b] \textbf{EnvTrace Accuracy (\%):} The percentage of test cases passing the strict binary criteria for semantic equivalence.\item[c] \textbf{Exact Match (\%):} The percentage of test cases where the generated code string is identical to a ground-truth solution \cite{mathur2024vision}.
        \item[d] \textbf{Normalized Levenshtein Distance (\%):} A value of $0\%$ indicates identical strings.
      \item[e] All results are the mean and standard deviation over three runs. \item[f] Models are grouped and then sorted by descending Average Full Score.
      \item[g] \textbf{No CodeBLEU on simple flows:} snippets are too short/non-self-contained for stable AST/data-flow scoring
      \item[t] Models run with reasoning enabled. ``(high)'' and ``(minimal)'' denote different reasoning levels.
    \end{tablenotes}
  \end{threeparttable}
\end{table}

\pgfplotstableread[col sep=semicolon]{
label;fullscore;exact;improvement;total
{Claude 3.5 Sonnet};99.5;89.7;6.9;96.6
{Grok-4};99.3;91.4;4.3;95.7
{Claude Opus 4 (Abacus)};99.1;89.7;6.0;95.7
{Devstral Medium};98.9;92.0;5.2;97.1
{Claude Sonnet 4 (Thinking)};98.9;91.7;4.0;95.7
{GPT-5 (high)};98.8;84.2;11.8;96.0
{GPT-4o};98.2;88.5;6.0;94.5
{Gemini 2.5 Pro};98.2;71.8;20.4;92.2
{Qwen3-Coder (480B)};97.8;90.5;5.7;96.3
{Athene-v2 (72.7B)};60.1;49.1;4.3;53.4
{Llama3.3 (70.6B)};52.4;35.3;4.3;39.7
{Qwen2.5 (7.62B)};50.7;33.6;5.2;38.8
{Mistral (7.3B)};20.7;14.4;1.7;16.1
{Phi-3.5 (3.8B)};0.4;0.3;0.0;0.3
}\datatable

\subsection{Model Performance on Complex-Flow Tasks}
\label{sec:complex_flow_results}  

\begin{figure}[ht]
\makebox[\textwidth][c]{%
\begin{minipage}{1.12\textwidth}
\centering
\caption{Model performance on complex-flow tasks. Left blue bars give the EnvTrace full score, while right stacked bars provide the string-based exact match (bottom, orange) and improvement (top, green) with EnvTrace accuracy.}
\label{fig:complex_tasks_performance}

\begin{tikzpicture}
\begin{axis}[
    width=1.05\textwidth, height=8cm,
    ybar,
    bar width=4pt,
    enlarge x limits=0.06,
    symbolic x coords={ {Beamline Scientist 1},{Claude Sonnet 4 (Thinking)},{Gemini 2.5 Pro},{Claude Opus 4 (Thinking)},{GPT-5 (high)},{Claude 3.5 Sonnet},{Grok-4},{GPT-4o},{Beamline Scientist 2},{Qwen3-Coder (480B)},{Devstral Medium},{Athene-v2 (72.7B)},{Llama3.3 (70.6B)},{Qwen2.5 (7.62B)},{Mistral (7.3B)},{Phi-3.5 (3.8B)} },
    xtick=data,
    xticklabel style={rotate=45,anchor=east,font=\scriptsize},
    ylabel={Performance (\%)},
    ymin=0, ymax=115,
    ymajorgrids=true,
    xmajorgrids=false,
    legend style={
        at={(0.5,1.02)}, anchor=south,
        legend columns=-1, /tikz/every even column/.append style={column sep=0.6cm},
        draw=none, fill=none, font=\small
    },
]

\addplot+[area legend, fill=oiBlue, bar shift=-2.5pt,
          nodes near coords,
          every node near coord/.append style={font=\scriptsize,rotate=90,anchor=west,inner sep=1pt,text=black},
          point meta=y]
    table[x=label,y=fullscore]{\datatable};
\addlegendentry{EnvTrace Full Score (\%)}

\addplot+[area legend, ybar, stack plots=y, fill=oiOrange, bar shift=+2.5pt]
    table[x=label,y=exact]{\datatable};
\addlegendentry{Exact Match (\%)}

\addplot+[ybar, stack plots=y, fill=oiBluishGreen, bar shift=+2.5pt, forget plot]
  table[x=label,y=improvement]{\datatable};

\addlegendimage{area legend, ybar, draw=none, shade,
  bottom color=oiOrange, top color=oiBluishGreen}
\addlegendentry{EnvTrace Accuracy (\%)}

\addplot+[draw=none, fill=none, bar shift=+2.5pt,
          nodes near coords, point meta=explicit,
          every node near coord/.append style={
            font=\scriptsize, rotate=90, anchor=west, inner sep=1pt, text=black
          }]
    table[x=label,y=total,meta expr=\thisrow{total}]{\datatable};

\end{axis}
\end{tikzpicture}
\end{minipage}%
}
\end{figure}
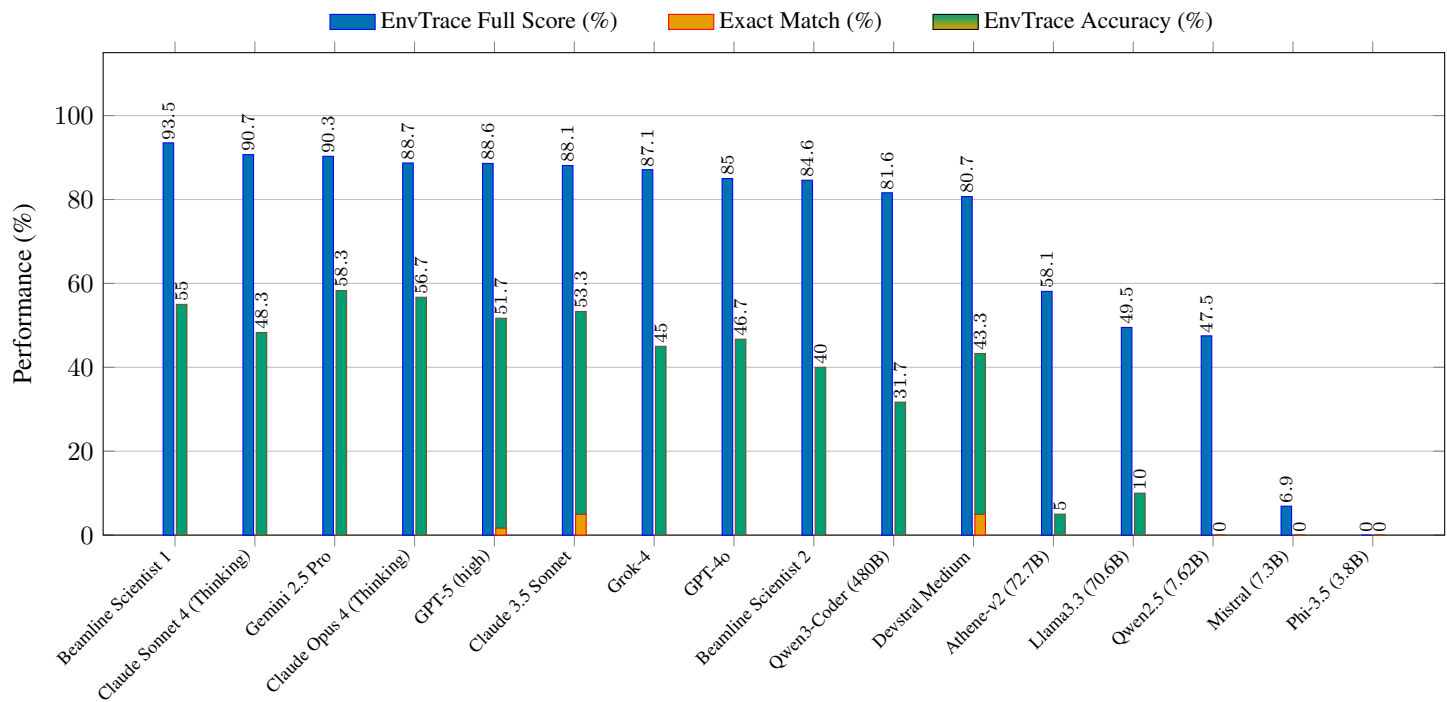

Tasks involving complex control flows, such as nested loops (\texttt{for} or \texttt{while}) and conditional statements (\texttt{if}), present greater challenges for both LLM reasoning and evaluation due to the increased ambiguity in natural language interpretation and the variability of possible implementations. The performance of 31 LLMs on the 20 complex-flow examples is provided in Table~\ref{tab:complex_results} and plotted in Fig.~\ref{fig:complex_tasks_performance}.
The full score and accuracy metrics offer different perspectives on how to interpret the evaluation.
For example, \texttt{Claude-Sonnet-4-(Thinking)} achieves a full score of 91\% but only 48\% in accuracy. 
This accuracy means that around half of the time the generated code was not perfect, regardless of the magnitude of the error/mismatch. The 91\% full score indicated that these mismatches were minor.
The breakdown of the full score into PV match, timing score, and temperature score is shown in Fig.~\ref{fig:complex_tasks_components} in Section~\ref{si:breakdown}. 
Accuracy and the degree of mismatch can both be valuable aspects of the evaluation.
The choice of metric can be adjusted depending on the specific application to better align with user needs.

The same top-tier closed-source models continue to lead for complex-flow tasks, with full score mostly above 85\% and accuracy around 35-55\%. 
These scores are comparable to those of human scientists, suggesting that LLMs can generate useful code drafts at human level, while final validation still requires human oversight or simulation-based evaluation.
The imperfect scoring largely reflects ambiguities in the natural language prompts rather than actual errors.
For example, after a series of measurements, one may choose to stay at the last measured position, return to the starting original position, or predict and move to a fresh sample spot to be ready for the next measurement. The order of the motor scanning can also introduce mismatch but do not result in actual error in the experiment. 
These varying interpretations are all valid, but differences in implementation lead to variations in the PV changes.
If scattering data were simulated, one could envision adjusting the metric to take the scattering data or features into account rather than solely relying on the PVs. 
Different models, influenced by their stochastic characteristics, exhibit varying behaviors.
For example, when asked to perform a map scan along sample x and y axes, most closed-source models would do a simple grid scan, while \texttt{GPT-5-minimal} and \texttt{high} (reasoning levels) would sometimes suggest a zigzag scan to minimize motor movement for efficient scanning, as shown in Box~\ref{lst:complex} with PV traces in Box~\ref{lst:PVcomplex}.
While all \texttt{Grok} models perform grid scan,
\texttt{Grok-Code-Fast-1} sometimes uses redundant motor movement: instead of moving sample x from 0.15 directly to 0.3, it moves from 0.15 back to 0 and then to 0.3. 
The averaged full score reflects the overall model performance and reliability, whereas the individual code comparison (per query) reflect specific errors or mismatches for easy bug detection or code improvement.
Most open-source models struggled with this map scan task, frequently producing hallucinated functions, confusing relative and absolute motor movements, or generating incomplete code. Among them, only the \texttt{Athene} models consistently produced correct executions for the grid scan case, although their overall performance remained limited. In contrast, \texttt{Qwen3-Coder-(480B)} performed significantly better across most scenarios, but failed once (out of three runs) in the grid scan due to an incorrect Python function call, resulting in a TypeError and thus a score of 0.
Aside from \texttt{Qwen3-coder-(480B)} that delivered comparable performance to closed-source models, in general the performance of open-source models did not exceed 60\% in full score and failed to achieve accuracy above 10\%, demonstrating that open-source models generally struggle with logic-based tasks. When provided with a much shorter prompt with limited beamline functionalities, performance of open-source models improves, as shown in Table~\ref{tab:complex_results_old_prompt}.

\begin{lstlisting}[float, language=Python, caption={Example of a complex-flow task and LLM code, showing that while closed-source model \texttt{GPT-5-high} provided efficient zigzag scan, open-source model use motor movement commands incorrectly.}, label={lst:complex}]
NL INPUT: Do a map scan, x range from 0 to 0.3mm, y from 0 to 0.6mm, step size is 0.15 horizontally and 0.2 vertically. (Exposure time 1s.)
-----------------------------------------------------
#GROUND-TRUTH SNIPPET:
for x in np.arange(0, 0.3 + 0.15/2, 0.15):
    for y in np.arange(0, 0.6 + 0.2/2, 0.2):
        sam.xabs(x)
        sam.yabs(y)
        sam.measure(1)
======================================================
#Gpt-5-high PREDICTED SNIPPET:
exposure_time = 1.0
step_x = 0.15
step_y = 0.2
x_end = 0.3
y_end = 0.6

nx = int(round(x_end / step_x)) + 1
ny = int(round(y_end / step_y)) + 1

for j in range(ny):
    for i in range(nx):
        sam.measure(exposure_time)
        if i < nx - 1:
            sam.xr(step_x if j % 2 == 0 else -step_x)
    if j < ny - 1:
        sam.yr(step_y)  
#-----------------------------------------------------
#PV match rate (non-temp): 60.00%
#Timing match: True (score: 0.871)
#EnvTrace Accuracy: False (Full score: 0.654)
======================================================
#Athene-v2 PREDICTED SNIPPET:
for x in np.arange(0, 0.3 + 0.15/2, 0.15):
    for y in np.arange(0, 0.6 + 0.2/2, 0.2):
        sam.xr(x)
        sam.yr(y)
        sam.measure(1)
#-----------------------------------------------------
#PV match rate (non-temp): 46.94%
#Timing match: False (score: 0.780)
#EnvTrace Accuracy: False (Full score: 0.531)
  
\end{lstlisting}

\pgfplotstableread[col sep=semicolon]{
label;fullscore;exact;improvement;total
{Beamline Scientist 1};93.5;0.0;55.0;55.0
{Claude Sonnet 4 (Thinking)};90.7;0.0;48.3;48.3
{Gemini 2.5 Pro};90.3;0.0;58.3;58.3
{Claude Opus 4 (Thinking)};88.7;0.0;56.7;56.7
{GPT-5 (high)};88.6;1.7;50.0;51.7
{Claude 3.5 Sonnet};88.1;5.0;48.3;53.3
{Grok-4};87.1;0.0;45.0;45.0
{GPT-4o};85.0;0.0;46.7;46.7
{Beamline Scientist 2};84.6;0.0;40.0;40.0
{Qwen3-Coder (480B)};81.6;0.0;31.7;31.7
{Devstral Medium};80.7;5.0;38.3;43.3
{Athene-v2 (72.7B)};58.1;0.0;5.0;5.0
{Llama3.3 (70.6B)};49.5;0.0;10.0;10.0
{Qwen2.5 (7.62B)};47.5;0.0;0.0;0.0
{Mistral (7.3B)};6.9;0.0;0.0;0.0
{Phi-3.5 (3.8B)};0.0;0.0;0.0;0.0
}\datatable

\pgfplotstablecreatecol[
  create col/expr={\thisrow{exact} + \thisrow{improvement}}
]{total}{\datatable}

\begin{table}[htbp]
  \centering
  \begin{threeparttable}
    \caption{Performance of LLMs on Complex-Flow Code Generation Tasks (N=20)}
    \label{tab:complex_results}
    \sisetup{separate-uncertainty=true, table-align-uncertainty=true}
    \begin{tabular}{
      l
      S[table-format=2.2(2.2)]
      S[table-format=2.2(2.2)]
      S[table-format=2.2(2.2)]
      S[table-format=3.2(2.2)]
      S[table-format=3.2(2.2)]
    }
      \toprule
      \textbf{Model} & 
      {\textbf{\begin{tabular}[c]{@{}c@{}}EnvTrace \\ Full Score (\%, $\uparrow$)\tnote{a}\end{tabular}}} & 
      {\textbf{\begin{tabular}[c]{@{}c@{}}EnvTrace \\ Accuracy (\%, $\uparrow$)\tnote{b}\end{tabular}}} & 
      \textcolor{gray}{\textbf{\begin{tabular}[c]{@{}c@{}}CodeBLEU \\ (comb-7, \%, $\uparrow$)\tnote{c}\end{tabular}}} &
      \textcolor{gray}{\textbf{\begin{tabular}[c]{@{}c@{}}Norm. Lev. \\ Dist. (\%, $\downarrow$)\tnote{d}\end{tabular}}} &
      {\textbf{\begin{tabular}[c]{@{}c@{}}Inference \\ Time (s)\end{tabular}}} \\
\midrule
\multicolumn{6}{l}{\textbf{Closed Source Models}} \\
\midrule
\multicolumn{6}{l}{\textit{Anthropic}} \\
  Claude Sonnet 4 (Thinking)\tnote{t} & 90.7 \pm 1.3 & 48.3 \pm 2.9 & 61.4 \pm 1.1 & 40.8 \pm 2.1 & 9.1 \pm 0.4 \\
  Claude Opus 4 (Thinking)\tnote{t} & 88.7 \pm 2.4 & 56.7 \pm 7.6 & 57.8 \pm 1.0 & 43.3 \pm 0.6 & 12.2 \pm 1.2 \\
  Claude Sonnet 4.5 (Thinking)\tnote{t} & 88.4 \pm 4.7 & 48.3 \pm 5.8 & 60.2 \pm 2.1 & 40.2 \pm 1.7 & 15.4 \pm 0.7 \\
  Claude 3.5 Sonnet & 88.1 \pm 0.5 & 53.3 \pm 2.9 & 57.2 \pm 0.8 & 39.1 \pm 1.1 & 3.1 \pm 0.1 \\
  Claude Opus 4 (Abacus) & 82.8 \pm 0.4 & 35.0 \pm 0.0 & 54.5 \pm 0.2 & 46.8 \pm 0.2 & 7.2 \pm 0.6 \\
  Claude Sonnet 4 (Abacus) & 80.4 \pm 0.0 & 35.0 \pm 0.0 & 57.0 \pm 0.0 & 42.0 \pm 0.0 & 4.2 \pm 0.4 \\
  Claude Sonnet 4 (Bedrock) & 80.0 \pm 0.2 & 33.3 \pm 2.9 & 57.7 \pm 0.2 & 41.9 \pm 0.5 & 11.2 \pm 2.1 \\
\midrule
\multicolumn{6}{l}{\textit{Google}} \\
  Gemini 2.5 Pro\tnote{t} & 90.3 \pm 0.6 & 58.3 \pm 2.9 & 60.3 \pm 2.0 & 39.6 \pm 0.9 & 16.2 \pm 0.9 \\
\midrule
\multicolumn{6}{l}{\textit{xAI}} \\
  Grok-4 Fast & 90.3 \pm 0.7 & 60.0 \pm 0.0 & 57.4 \pm 1.8 & 47.4 \pm 1.0 & 10.2 \pm 0.7 \\
  Grok-4\tnote{t} & 87.1 \pm 1.1 & 45.0 \pm 0.0 & 55.9 \pm 2.1 & 44.3 \pm 3.8 & 74.8 \pm 7.2 \\
  Grok Code Fast 1 & 85.2 \pm 2.3 & 46.7 \pm 2.9 & 58.4 \pm 2.4 & 44.5 \pm 1.3 & 7.4 \pm 1.5 \\
\midrule
\multicolumn{6}{l}{\textit{OpenAI / Microsoft}} \\
  GPT-5 (high)\tnote{t} & 88.6 \pm 2.1 & 51.7 \pm 7.6 & 53.0 \pm 2.8 & 50.4 \pm 2.7 & 108.2 \pm 12.9 \\
  GPT-4o & 85.0 \pm 1.2 & 46.7 \pm 5.8 & 54.3 \pm 0.4 & 46.1 \pm 1.2 & 1.8 \pm 0.0 \\
  o3 (high)\tnote{t} & 84.7 \pm 1.6 & 45.0 \pm 8.7 & 56.7 \pm 1.1 & 42.9 \pm 1.5 & 15.7 \pm 0.4 \\
  GPT-5 (minimal)\tnote{t} & 83.9 \pm 3.8 & 33.3 \pm 7.6 & 49.8 \pm 1.3 & 51.7 \pm 1.2 & 4.1 \pm 0.4 \\
  GPT-4o (Abacus) & 83.7 \pm 1.1 & 48.3 \pm 5.8 & 54.4 \pm 0.6 & 44.2 \pm 0.9 & 3.5 \pm 0.3 \\
\midrule
\multicolumn{6}{l}{\textit{Mistral}} \\
  Devstral Medium\tnote{t} & 80.7 \pm 0.6 & 43.3 \pm 5.8 & 56.9 \pm 0.4 & 40.9 \pm 1.4 & 5.0 \pm 4.3 \\
\midrule
\multicolumn{6}{l}{\textbf{Open Source Models}} \\
\midrule
  Qwen3-Coder (480B) & 81.6 \pm 2.5 & 31.7 \pm 2.9 & 49.9 \pm 0.8 & 46.3 \pm 1.4 & 4.8 \pm 0.3 \\
  Athene-v2 (72.7B) & 58.1 \pm 0.0 & 5.0 \pm 0.0 & 57.2 \pm 0.0 & 44.5 \pm 0.0 & 8.8 \pm 0.7 \\
  Qwen3-Coder (30.5B) & 57.4 \pm 1.1 & 5.0 \pm 0.0 & 46.5 \pm 0.1 & 51.1 \pm 0.4 & 3.0 \pm 0.2 \\
  Athene-v2-Agent (72.7B) & 53.3 \pm 0.2 & 0.0 \pm 0.0 & 55.8 \pm 0.2 & 43.9 \pm 0.7 & 8.0 \pm 0.6 \\
  Llama3.3 (70.6B) & 49.5 \pm 1.1 & 10.0 \pm 0.0 & 49.9 \pm 0.4 & 49.4 \pm 0.2 & 9.3 \pm 0.7 \\
  Qwen2.5 (7.62B) & 47.5 \pm 0.1 & 0.0 \pm 0.0 & 47.2 \pm 2.0 & 48.5 \pm 1.0 & 1.8 \pm 0.5 \\
  Qwen2.5-Coder (32.8B) & 42.7 \pm 0.6 & 5.0 \pm 0.0 & 57.0 \pm 0.0 & 40.8 \pm 0.1 & 4.7 \pm 0.2 \\
  Mistral-NeMo (12.2B) & 33.6 \pm 1.6 & 5.0 \pm 0.0 & 52.2 \pm 1.2 & 46.3 \pm 0.5 & 2.2 \pm 0.4 \\
  GPT-oss (117B) & 32.4 \pm 0.0 & 5.0 \pm 0.0 & 48.5 \pm 0.0 & 72.2 \pm 0.0 & 213.0 \pm 0.6 \\
  Qwen2 (7.62B) & 25.9 \pm 0.5 & 5.0 \pm 0.0 & 45.8 \pm 0.3 & 59.9 \pm 0.5 & 1.6 \pm 0.2 \\
  Mistral (7.3B) & 6.9 \pm 0.0 & 0.0 \pm 0.0 & 39.6 \pm 0.0 & 63.0 \pm 0.0 & 1.7 \pm 0.0 \\
  Phi-3.5 (3.8B) & 0.0 \pm 0.0 & 0.0 \pm 0.0 & 40.0 \pm 0.0 & 100.0 \pm 0.0 & 0.7 \pm 0.0 \\
  Qwen3 (32B) & 0.0 \pm 0.0 & 0.0 \pm 0.0 & 38.4 \pm 0.0 & 89.2 \pm 0.0 & 14.0 \pm 0.1 \\
  Phi-3.5 (3.8B, fp16) & 0.0 \pm 0.0 & 0.0 \pm 0.0 & 40.0 \pm 0.0 & 100.0 \pm 0.0 & 0.6 \pm 0.0 \\
\midrule
\multicolumn{6}{l}{\textbf{Humans}} \\
\midrule
  Beamline Scientist 1 & 93.5 \pm 0.0 & 55.0 \pm 0.0 & 54.1 \pm 0.0 & 44.8 \pm 0.0 & \multicolumn{1}{c}{—} \\
  Beamline Scientist 2 & 84.6 \pm 0.0 & 40.0 \pm 0.0 & 48.6 \pm 0.0 & 42.8 \pm 0.0 & \multicolumn{1}{c}{—} \\
\bottomrule
    \end{tabular}
    \begin{tablenotes}
      \small
      \item[a] \textbf{EnvTrace Full Score (\%):} The continuous score from 0 to 100, reflecting the weighted average of state and temporal fidelity.
      \item[b] \textbf{EnvTrace Accuracy (\%):} The percentage of test cases passing the strict binary criteria for semantic equivalence.
      \item[c] \textbf{CodeBLEU (comb-7, \%):} syntactic similarity metric. We report the ``comb-7'' variant, which weights data-flow and syntax more heavily \cite{ren2020codebleu}.
      \item[d] \textbf{Normalized Levenshtein Distance (\%):} A value of $0\%$ indicates identical strings.
      \item[e] All results are the mean and standard deviation over three runs. \item[f] Models are grouped and then sorted by descending Average Full Score.
      \item[h] \textbf{Human Inference Time} in the order of minutes (>100s)
      \item[t] Models run with reasoning enabled. ``(high)'' and ``(minimal)'' denote different reasoning levels.
    \end{tablenotes}
  \end{threeparttable}
\end{table}


\subsection{Discover Alternative Plans}
The EnvTrace method can be readily adapted to any codebase or beamline for evaluating code and evaluating functional equivalence. We demonstrate this using native Bluesky plans~\cite{bluesky}, which are a set of several dozen pre-defined experimental procedures that serve as fundamental building blocks for constructing custom beamline operations.
As shown in Table~\ref{tab:simple_bluesky_results} and Fig.~\ref{fig:bluesky_performance}, even with minimal documentation available online, LLMs can effectively generate instrument-control code, and when paired with EnvTrace, their outputs can be meaningfully evaluated. 
For example, with the natural language query ``Do a map scan over smx and smy, x from 0 to 5 over 6 steps, y from 0 to 10 with 5 steps'', 
\texttt{GPT-4o} used a different function with comments, as shown in Box~\ref{lst:bluesky}.
Exact string or syntatic metrics would fail entirely in this case. However, EnvTrace returned a perfect PV match for all 94 PV changes, with slight mismatch on timing due to simulator timing variation, producing a full score of 97\%.
Open-source models like \texttt{Llamma 3.3} and \texttt{Athene} tend to make minor but fatal edits to the commands, e.g., incorrect input, variable type, or name.
Bluesky plans match simple-flow cases in brevity but differ in that they encapsulate a more intricate sequence of actions.
Each Bluesky plan triggers a series of actions with several PV changes, showing that EnvTrace with a simulator is especially important to track simple code with internal procedures.

\begin{lstlisting}[float, language=Python, caption={Example of LLM generating a funtionally equivalent Bluesky code}, label={lst:bluesky}] 
NL INPUT: Do a map scan over smx and smy, x from 0 to 5 over 6 steps, y from 0 to 10 with 5 steps
------------------------------------------------------
#GROUND-TRUTH SNIPPET:
RE(outer_product_scan(cms.detector, smx , 0, 5, 6, smy , 0, 10, 5))
======================================================
#GPT-4o PREDICTED SNIPPET: 
# Perform a grid scan over smx and smy, with smx from 0 to 5 in 6 steps and smy from 0 to 10 in 5 steps
RE(grid_scan(cms.detector, smx, 0, 5, 6, smy, 0, 10, 5))
#-----------------------------------------------------
#PV match rate (non-temp): 100.00%
#Timing match: True (score: 0.872)
#Full match: True (score: 0.974)
\end{lstlisting}


\section{Discussion}
\label{sec:discussion}

We have demonstrated three categories of LLM-generated code evaluated with EnvTrace: (1) simple-flow code involving minimal instrument changes, (2) complex-flow incorporating control logic that allows creative solutions, and (3) simple code with internal procedures where LLMs can produce valid but different implementations.
LLMs can generate codes with stylistic variation, creative logic, or alternative implementations;
EnvTrace provides a robust and generalizable framework for evaluating LLMs in instrument-control settings by focusing on semantic, execution-based assessment rather than brittle syntactic metrics. 
This approach supports safer deployment, ensuring that only functionally correct code is executed on costly and sensitive instruments. It enables cross-beamline benchmarking, offering a unified evaluation method that can be applied consistently across different beamlines and facilities, thereby making performance results transferable and comparable. 
Below, we summarize the key findings and discuss future directions.

\subsection{Semantic over Syntactic}
\label{sec:syntactic_vs_semantic}

A central claim of our work is that static syntactic-based code metrics do not reliably reflect the functional correctness essential for physical systems. 
Evaluating with functional correctness allows for the identification and correction of two critical failure modes of syntactic metrics, as discussed below.

\textbf{Identify False Negatives:} High-performing models like \texttt{Claude-3.5-Sonnet} achieve excellent EnvTrace full scores (over 80\%) but are heavily penalized by syntactic metrics (CodeBLEU score around 50\%). This occurs when the model generates a valid, alternative implementation that is stylistically different from the ground truth. EnvTrace correctly identifies this code as successful, whereas a purely syntactic evaluation would incorrectly flag it as low-quality.

\textbf{Identify False Positives:} Conversely, models such as \texttt{Athene-v2} produce code that appears syntactically plausible (CodeBLEU score approx. 60\%) but contains critical semantic errors, resulting in a lower functional score. These subtle but critical bugs, such as incorrect function name (as shown in Box~\ref{lst:simple}) or incorrect parameter in a function call, can cause minor variations in syntactic metrics but are immediately caught by our execution-based framework. 

To further investigate these, we examined the relationship between our semantic EnvTrace full score and two widely-used syntactic metrics, CodeBLEU and normalized Levenshtein distance (nLD), with details provided in Section~\ref{si:relationship_analysis}.
Figure~\ref{fig:si_levenshtein_simple} shows the comparison between full score and normalized Levenshtein distance (nLD) for simple-flows.
As longer Levenshtein distance indicates lower performance, here the higher EnvTrace full score roughly correlates with higher performance determined by nLD.
While there is a correlation for simple-flows, many LLMs with an nLD between 20--40\% exhibit significant variation in full score, ranging from around 20--80\%. 
This shows that, despite the overall correlation, the full score provides a metric that more accurately reflects the actual performance.
As shown by the \texttt{Qwen2.5-coder} example in Box~\ref{lst:simple}, when part of the generated code provides correct execution, full score is able to capture the partial success.
For complex-flows, comparison of model performance with full score versus CodeBLEU is provided in Fig.~\ref{fig:si_codebleu_complex}.
For most LLMs, CodeBLEU scores are clustered around 40--60\%, while the full score varies from 0--90\%, rendering the CodeBLEU a weak indicator for performance.
As shown by the full score versus nLD in Fig.~ \ref{fig:si_levenshtein_complex}, 
except for very weak models that give higher nLD, the nLD metric places most LLMs around 40--60\%, while again the full scores give a full range of around 0--90\%.
The lack of correlation between full score and syntactic metrics is expected as any slight variation in the variable name or the use of different but equivalent logic would be heavily penalized by syntactic methods.
Semantic and syntactic differences are most significant in complex-flow problems or sequence of actions, while syntactic similarity is a better, yet imperfect, indicator for simple tasks.
This growing divergence with task complexity underscores the necessity of semantic execution-based evaluation for developing robust and reliable real-world AI agents.
The choice of metric can be tailored to the system by adjusting parameters based on the simulator’s capabilities and experiment complexity, such as choosing between full score or accuracy and tuning the sensitivity of each component to suit the system.

\subsection{Interpreting Functional Performance}

Our results clearly demonstrate that a semantic, execution-based evaluation framework like EnvTrace, provides a meaningful and insightful assessment of code-generating agents than traditional syntactic metrics. 
The continuous full score provides a practical metric for evaluating the quality of generated code, but its numerical value must be interpreted in the context of real-world beamline operations. Based on our human expert review of the execution traces, we have developed a heuristic for translating EnvTrace scores into practical assessments. A score above \textbf{90\%} typically indicates a robust and reliable solution, where any minor deviations are functionally inconsequential. Scores between \textbf{70-90\%} often represent code that is largely correct but may have minor flaws in timing or non-critical parameters. Scores below \textbf{60\%}, however, almost always indicate a significant functional failure where the primary goal of the command was not achieved. This interpretive framework, validated by beamline scientists, allows the full score to serve as a reliable guide for both model development and operational decision-making.
Small variations in the score, typically a few percents, can arise from the inherent stochasticity of the model or the simulated physical system, such as minor overheads in motor or detector responses that affect the timing component. This reflects the reality of working with physical hardware and underscores the value of a metric that is robust to such real-world fluctuations.

A key challenge in this domain is the nature of the `ground truth' itself. Many natural language commands are inherently ambiguous and can have multiple valid implementations. For this reason, our ground-truth dataset should be viewed as a collection of `expected answers' rather than an exclusive, rigid truth. A deviation from a ground-truth solution is not necessarily an error, but rather a ‘mismatch’ that may represent a valid alternative implementation.
This nuance is clearly reflected in our human benchmark results. Even expert-written code did not achieve a perfect score right away, due to a combination of minor typos or bugs, different interpretations of the ambiguous natural language queries, or valid procedural differences. 
For instance, after a scan, one scientist might program the sample to return to its starting position while another might leave it at the end position; one person may interpret measurement interval as inclusive or exclusive of measurement time itself. 
If provided with a comprehensive list of possible GTs and carefully debugged human-generated code, the human performance presented in Section~\ref{sec:complex_flow_results} would achieve perfect results.
However, instead of generating this idealized performance, the human performance we presented here is a realistic benchmark based on quick responses from human scientists and non-exclusive GTs, i.e. multiple valid answers may exist beyond those represented in our evaluation.
Therefore, the competitiveness of top-tier LLMs with the human benchmark strongly demonstrates their capacity to accelerate code generation.

For LLMs to be effectively integrated into workflows, it is essential to first establish basic infrastructure, such as clear documentation, well-structured code, and organized data and metadata. These steps are relatively straightforward and not technically demanding, yet they are crucial for making instruments and datasets `AI-ready'. Once in place, this foundation enables LLMs to automate tasks easily and reliably, paving the way for intelligent agentic workflows.

\subsection{Symbiotic LLM and Digital Twin Systems}

\begin{figure}[b]
  \centering
  \includegraphics[width=\textwidth]{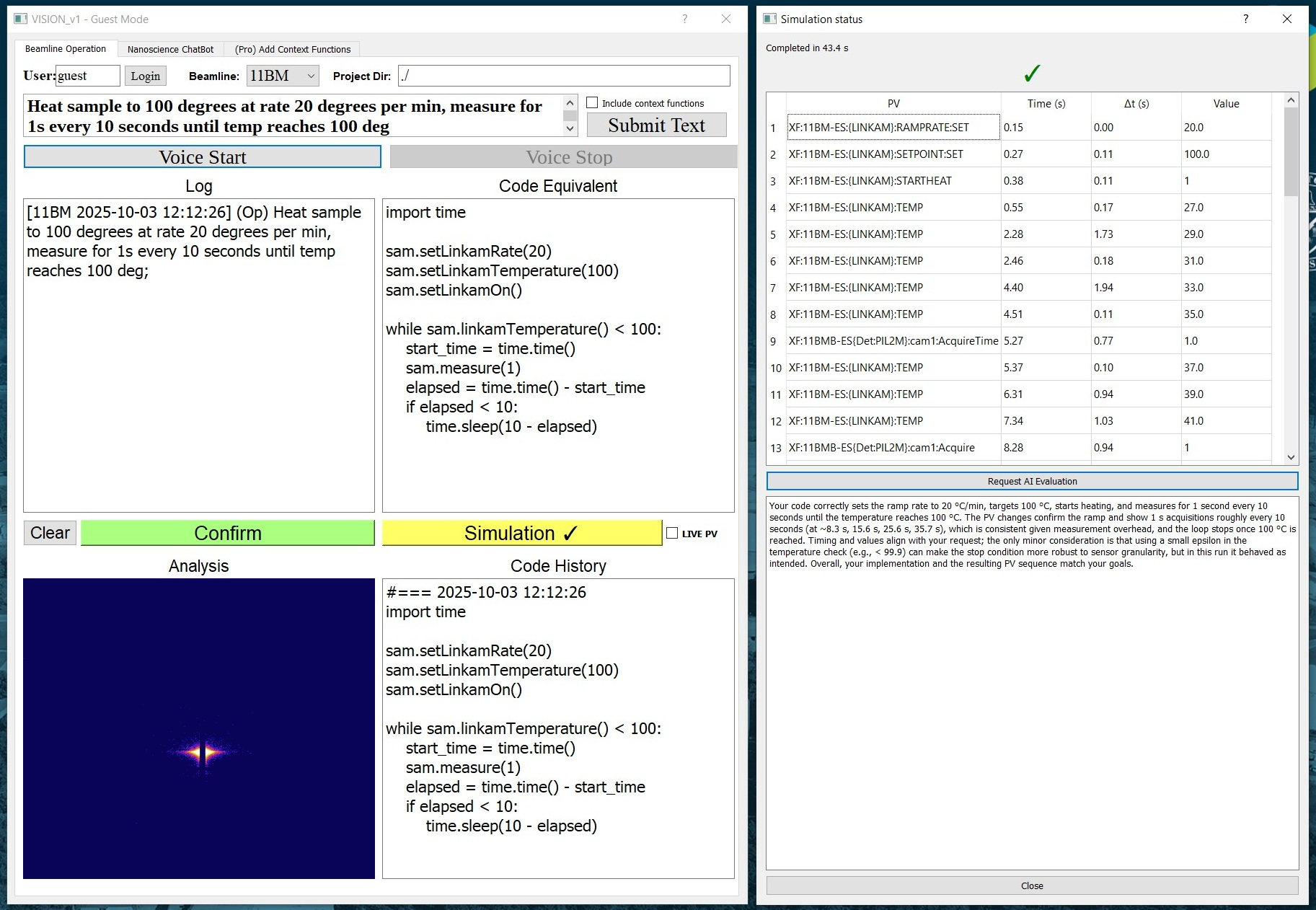}
  \caption{Illustration of the VISION GUI showing an example natural language query for performing in-situ thermal measurements, along with the corresponding LLM-generated code, partial PV traces, and an AI evaluation based on these inputs (query, code, and PV traces). 
  }
  \label{fig:gui}
\end{figure}

The simulator serves not only as part of AI evaluation but also as an integral component of the experimental workflow for human users. In initial attempts, human-written code often contains minor syntax errors or typos, particularly during time-sensitive stressful beamtime sessions. The coding capabilities of LLMs provide rapid code drafts for human review, while the simulator environment offers runtime feedback that enables fast, iterative debugging. This process can occur in real time during experiments or as a `pre-flight check' for new experimental procedures, eliminating the need for direct beamline access and beamtime consumption. 
We have integrated the simulator into VISION~\cite{mathur2024vision} as part of the Operator for real-time operation, as shown in Fig.~\ref{fig:gui}.
This integration supports infrastructure-aware generation by loading the beamline-specific configuration files, ensuring that an LLM will generate code for the specific features and capabilities of the beamline. 
Users can interact with VISION using natural language, after which an LLM generates the corresponding code; users can then review, edit, and simulate the code as needed, providing a flexible and intuitive interface for controlling and testing instrument code.
With simulation, PV traces from successful execution and errors (if any) from the IPython session will be displayed. Once the simulation is complete, the user can choose to get an AI evaluation, as described in Section~\ref{si:ai}.
This AI evaluation takes in the natural language query, generated (or edited) code, and the PV changes stemming from the simulation, as shown by examples in Fig.~\ref{fig:gui} and Section~\ref{si:fig:square_example}.
To provide physical grounding of LLM-generated code, the beamline EPICS archiver is used to load the most recent instrument state and settings (PV values) into the simulator, enabling accurate and realistic predictions.
For improved beamline integration, developing a digital twin within the Bluesky ecosystem, e.g. Bluesky Adaptive, QueueServer, and Tiled, can facilitate more efficient and streamlined workflows.

Here, we used EnvTrace with a control-logic digital twin to track instrument changes. If an advanced simulator or digital twin is available, one could envision defining a metric based on the simulated scattering data~\cite{chen2025agentic} or features rather than solely relying on the instrument changes. 
Digital twins~\cite{yang2025leveraging, zhang2024large} are important for advancing physical systems by providing real-time, high-fidelity simulations that mirror scientific instruments or complex environments in scientific facilities~\cite{es-haghiMethodsEnablingRealtime2024, iranshahiDigitalTwinsRecent2025}. 
Different methodologies, instrument configurations, and experiments can be explored without constraints on instrument availability and physical location.   
LLMs and digital twins are mutually beneficial: incorporating LLMs into digital twins allows AI agents to interpret natural-language instructions and manage simulated and physical interactions, while digital twins, in turn, provide realistic environments for evaluating LLM performance.
Together, digital twins powered by LLMs within multi-agent ecosystems form the basis of embodied AI for enabling autonomous agents to perceive, reason, and act intelligently in real-world environments through complex, context-aware decision-making~\cite{kyager2024exocortex}.

\section{Conclusion}
\label{sec:conclusion}

Deploying LLM coding agents in physical systems is challenging due to the high risks involved with hardware interaction and the lack of reliable evaluation metrics.
EnvTrace addresses this by executing ground-truth and LLM-generated code in a simulator, comparing their execution traces to produce a multi-faceted score that captures correctness, timing, and continuous variable progression.
We have demonstrated three types of examples for LLM code generation evaluated with EnvTrace: (1) simple-flow code involving minimal instrument changes, (2) complex-flow incorporating control logic that allows creative solutions, and (3) simple code with internal sequences where LLMs can produce valid but different implementations.
LLMs can generate code with stylistic variation, creative logic, or alternative implementations.
EnvTrace provides a robust and generalizable framework for evaluating LLMs in instrument-control settings by focusing on interpretable semantic and execution-based assessment, rather than brittle syntactic metrics.

We have also demonstrated the integration of the EnvTrace simulator into a real-time assistant to support LLM-embedded beamline operation.
While our demonstrations are already deployable at the beamline, achieving broader impact and wider applicability will require further integration of LLMs within the Bluesky ecosystem, together with continued progress toward a full digital twin with e.g. simulated optics and X-ray data.
Effective evaluation of LLM coding agents require digital twins, while LLM-augmented digital twins enable more intelligent, adaptive control, highlighting the mutual benefits of their integration.
The integration of LLMs with digital twins not only advances the capabilities of physical AI but also establishes a foundation for embodied AI, enabling autonomous agents to learn, adapt, and operate reliably within complex real-world environments.


\section*{Acknowledgments}
The work was supported by a DOE Early Career Research Program.
This research also used beamline 11-BM (CMS) of the National Synchrotron Light Source II (NSLS-II) and utilized the X-ray scattering partner user program at the Center for Functional Nanomaterials (CFN), both of which are U.S. Department of Energy (DOE) Office of Science User Facilities operated for the DOE Office of Science by Brookhaven National Laboratory under Contract No. DE-SC0012704. 
We thank beamline scientists Ruipeng Li and Siyu Wu and data scientists Jennefer Maldonado, Muzamil Hussain Syed, and technology architect Kunal Shroff for their assistance on the project.

\bibliographystyle{unsrtnat}  
\bibliography{ref_envtrace}

@article{prince2023opportunities,
  title={Opportunities for retrieval and tool augmented large language models in scientific facilities},
  author={Prince, Michael H and Chan, Henry and Vriza, Aikaterini and Zhou, Tao and Sastry, Varuni K and Luo, Yanqi and Dearing, Matthew T and Harder, Ross J and Vasudevan, Rama K and Cherukara, Mathew J},
  journal={npj Computational Materials},
  volume={10},
  number={1},
  pages={251},
  year={2024},
  publisher={Nature Publishing Group UK London}
}

@article{kyager2024exocortex,
author ="Yager, Kevin G.",
title  ="{Towards a Science Exocortex}",
journal  ="Digital Discovery",
year  ="2024",
volume  ="3",
issue  ="10",
pages  ="1933-1957",
publisher  ="RSC",
doi  ="10.1039/D4DD00178H",
url  ="http://dx.doi.org/10.1039/D4DD00178H"}

@article{tsai2023vision,
author ="Daniel Potemkin and Carlos Soto and Ruipeng Li and Kevin Yager and Esther Tsai",
title  ="Virtual Scientific Companion for Synchrotron Beamlines: A Prototype",
journal  ="arxiv",
year  ="2023",
doi  ="10.48550/arXiv.2312.17180",
url  ="https://doi.org/10.48550/arXiv.2312.17180"}

@article{mathur2024vision,
  title={VISION: A modular AI assistant for natural human-instrument interaction at scientific user facilities},
  author={Mathur, Shray and van der Vleuten, Noah and Yager, Kevin G and Tsai, Esther},
  journal={Machine Learning: Science and Technology},
  year={2025}
}

@article{holler2017high,
  title={High-resolution non-destructive three-dimensional imaging of integrated circuits},
  author={Holler, Mirko and Guizar-Sicairos, Manuel and Tsai, Esther HR and Dinapoli, Roberto and M{\"u}ller, Elisabeth and Bunk, Oliver and Raabe, J{\"o}rg and Aeppli, Gabriel},
  journal={Nature},
  volume={543},
  number={7645},
  pages={402--406},
  year={2017},
  publisher={Nature Publishing Group}
}

@article{shahmoradian2017three,
  title={Three-dimensional imaging of biological tissue by cryo x-ray ptychography},
  author={Shahmoradian, Sarah H. and Tsai, Esther H. R. and Diaz, Ana and Guizar-Sicairos, Manuel and Raabe, Jorg and Spycher, L and Britschgi, M and Ruf, A and Stahlberg, H and Holler, Mirko},
  journal={Scientific reports},
  volume={7},
  number={1},
  pages={1--12},
  year={2017},
  publisher={Nature Publishing Group}
}

@article{sidhik2024two,
  title={Two-dimensional perovskite templates for durable, efficient formamidinium perovskite solar cells},
  author={Sidhik, Siraj and Metcalf, Isaac and Li, Wenbin and Kodalle, Tim and Dolan, Connor J and Khalili, Mohammad and Hou, Jin and Mandani, Faiz and Torma, Andrew and Zhang, Hao and others},
  journal={Science},
  volume={384},
  number={6701},
  pages={1227--1235},
  year={2024},
  publisher={American Association for the Advancement of Science}
}

@article{bluesky,
  author    = {Allan, Daniel and Caswell, Thomas and Campbell, Stuart and Rakitin, Maksim},
  journal   = {Synchrotron Radiation News},
  title     = {{Bluesky's Ahead: A Multi-Facility Collaboration for an a la Carte Software Project for Data Acquisition and Management}},
  year      = {2019},
  number    = {3},
  pages     = {19--22},
  volume    = {32},
  doi       = {10.1080/08940886.2019.1608121},
  publisher = {Taylor \& Francis},
  url       = {https://doi.org/10.1080/08940886.2019.1608121},
}

@article{holler2019three,
  title={Three-dimensional imaging of integrated circuits with macro-to nanoscale zoom},
  author={Holler, Mirko and Odstrcil, Michal and Guizar-Sicairos, Manuel and Lebugle, Maxime and M{\"u}ller, Elisabeth and Finizio, Simone and Tinti, Gemma and David, Christian and Zusman, Joshua and Unglaub, Walter and others},
  journal={Nature Electronics},
  volume={2},
  number={10},
  pages={464--470},
  year={2019},
  publisher={Nature Publishing Group UK London}
}

@article{aidukas2024high,
  title={High-performance 4-nm-resolution X-ray tomography using burst ptychography},
  author={Aidukas, Tomas and Phillips, Nicholas W and Diaz, Ana and Poghosyan, Emiliya and M{\"u}ller, Elisabeth and Levi, Anthony FJ and Aeppli, Gabriel and Guizar-Sicairos, Manuel and Holler, Mirko},
  journal={Nature},
  volume={632},
  number={8023},
  pages={81--88},
  year={2024},
  publisher={Nature Publishing Group UK London}
}

@book{willmott2019introduction,
  title={An introduction to synchrotron radiation: techniques and applications},
  author={Willmott, Philip},
  year={2019},
  publisher={John Wiley \& Sons}
}

@article{naik2024limitations,
  title={On the limitations of embedding based methods for measuring functional correctness for code generation},
  author={Naik, Atharva},
  journal={arXiv preprint arXiv:2405.01580},
  year={2024}
}

@article{chen2021evaluating,
  title={Evaluating large language models trained on code},
  author={Chen, Mark and Tworek, Jerry and Jun, Heewoo and Yuan, Qiming and Pinto, Henrique Ponde De Oliveira and Kaplan, Jared and Edwards, Harri and Burda, Yuri and Joseph, Nicholas and Brockman, Greg and others},
  journal={arXiv preprint arXiv:2107.03374},
  year={2021}
}

@inproceedings{papineni2002bleu,
  title={Bleu: a method for automatic evaluation of machine translation},
  author={Papineni, Kishore and Roukos, Salim and Ward, Todd and Zhu, Wei-Jing},
  booktitle={Proceedings of the 40th annual meeting of the Association for Computational Linguistics},
  pages={311--318},
  year={2002}
}

@article{ren2020codebleu,
  title={Codebleu: a method for automatic evaluation of code synthesis},
  author={Ren, Shuo and Guo, Daya and Lu, Shuai and Zhou, Long and Liu, Shujie and Tang, Duyu and Sundaresan, Neel and Zhou, Ming and Blanco, Ambrosio and Ma, Shuai},
  journal={arXiv preprint arXiv:2009.10297},
  year={2020}
}

@article{jiang2024survey,
  title={A survey on large language models for code generation},
  author={Jiang, Juyong and Wang, Fan and Shen, Jiasi and Kim, Sungju and Kim, Sunghun},
  journal={arXiv preprint arXiv:2406.00515},
  year={2024}
}

@article{tong2024codejudge,
  title={Codejudge: Evaluating code generation with large language models},
  author={Tong, Weixi and Zhang, Tianyi},
  journal={arXiv preprint arXiv:2410.02184},
  year={2024}
}

@article{zhuo2023ice,
  title={Ice-score: Instructing large language models to evaluate code},
  author={Zhuo, Terry Yue},
  journal={arXiv preprint arXiv:2304.14317},
  year={2023}
}

@article{zhou2023codebertscore,
  title={Codebertscore: Evaluating code generation with pretrained models of code},
  author={Zhou, Shuyan and Alon, Uri and Agarwal, Sumit and Neubig, Graham},
  journal={arXiv preprint arXiv:2302.05527},
  year={2023}
}

@article{kulal2019spoc,
  title={Spoc: Search-based pseudocode to code},
  author={Kulal, Sumith and Pasupat, Panupong and Chandra, Kartik and Lee, Mina and Padon, Oded and Aiken, Alex and Liang, Percy S},
  journal={Advances in Neural Information Processing Systems},
  volume={32},
  year={2019}
}

@article{chen2025agentic,
  title={An Agentic Artificially Intelligent X-ray Scientist},
  author={Chen, Zhantao and Petsch, Alexander and Israelski, Aidan and Plumley, Rajan and Shen, Lingjia and Wang, Cong and Peng, Cheng and Ni, Yuan and Bansil, Arun and Chowdhury, Sugata and others},
  year={2025}
}

@article{hellert2025agentic,
  title={Agentic AI for Multi-Stage Physics Experiments at a Large-Scale User Facility Particle Accelerator},
  author={Hellert, Thorsten and Bertwistle, Drew and Leemann, Simon C and Sulc, Antonin and Venturini, Marco},
  journal={arXiv preprint arXiv:2509.17255},
  year={2025}
}

@article{hellert2025alpha,
  title={Alpha Berkeley: A Scalable Framework for the Orchestration of Agentic Systems},
  author={Hellert, Thorsten and Montenegro, Jo{\~a}o and Sulc, Antonin},
  journal={arXiv preprint arXiv:2508.15066},
  year={2025}
}

@article{sulc2024towards,
  title={Towards agentic ai on particle accelerators},
  author={Sulc, Antonin and Hellert, Thorsten and Kammering, Raimund and Hoschouer, Hayden and John, Jason St},
  journal={arXiv preprint arXiv:2409.06336},
  year={2024}
}

@article{yao2025operationalizing,
  title={Operationalizing Serendipity: Multi-Agent AI Workflows for Enhanced Materials Characterization with Theory-in-the-Loop},
  author={Yao, Lance and Samantray, Suman and Ghosh, Ayana and Roccapriore, Kevin and Kovarik, Libor and Allec, Sarah and Ziatdinov, Maxim},
  journal={arXiv preprint arXiv:2508.06569},
  year={2025}
}

@article{rasheed2020digital,
  title={Digital twin: Values, challenges and enablers from a modeling perspective},
  author={Rasheed, Adil and San, Omer and Kvamsdal, Trond},
  journal={IEEE access},
  volume={8},
  pages={21980--22012},
  year={2020},
  publisher={IEEE}
}

@article{stadtmann2024physics,
  title={Physics-guided federated learning as an enabler for digital twins},
  author={Stadtmann, Florian and Furevik, Erik Rugaard and Rasheed, Adil and Kvamsdal, Trond},
  journal={Expert Systems with Applications},
  volume={258},
  pages={125169},
  year={2024},
  publisher={Elsevier}
}

@article{van2022executable,
  title={The executable digital twin: merging the digital and the physics worlds},
  author={Van der Auweraer, Herman and Hartmann, Dirk},
  journal={arXiv preprint arXiv:2210.17402},
  year={2022}
}

@inproceedings{dalesio2020epics,
  title={The EPICS Collaboration Turns 30},
  author={Dalesio, Leo and Johnson, Andrew and Kasemir, Kay-Uwe and others},
  booktitle={17th International Conference on Accelerator and Large Experimental Physics Control Systems (ICALEPCS'19), New York, NY, USA, 05-11 October 2019},
  pages={101--105},
  year={2020},
  organization={JACOW Publishing, Geneva, Switzerland}
}

@article{xiao2025application,
  title={Application of Synchrotron Radiation in Fundamental Research and Clinical Medicine},
  author={Xiao, Chao and Zhang, Jinde and Li, Yang and Xie, Mingyuan and Sun, Dongbai},
  journal={Biomedicines},
  volume={13},
  number={6},
  pages={1419},
  year={2025},
  publisher={MDPI}
}

@article{xu2024unraveling,
  title={Unraveling the Formation Mechanisms of Highly Oriented Tin Perovskite with a 3D-over-2D Heterostructure},
  author={Xu, Yuanze and Kim, Juno and Ramakrishnan, Shripathi and Chen, Taiyi and Zhang, Xiaoyu and Zhang, Yugang and Musser, Andrew J and Yu, Qiuming},
  journal={ACS Energy Letters},
  volume={9},
  number={9},
  pages={4734--4745},
  year={2024},
  publisher={ACS Publications}
}

@article{theeventhorizontelescopecollaborationFirstM87Event2019,
  title = {First {{M87 Event Horizon Telescope Results}}. {{I}}. {{The Shadow}} of the {{Supermassive Black Hole}}},
    author = {The Event Horizon Telescope Collaboration and Akiyama, Kazunori and Alberdi, Antxon and Alef, Walter and Asada, Keiichi and Azulay, Rebecca and others},
  year = {2019},
  month = apr,
  journal = {The Astrophysical Journal Letters},
  volume = {875},
  number = {1},
  pages = {L1},
  issn = {2041-8205, 2041-8213},
  doi = {10.3847/2041-8213/ab0ec7},
  urldate = {2025-09-16},
  langid = {english}
}

@article{hendrycksapps2021,
  title={Measuring Coding Challenge Competence With APPS},
  author={Dan Hendrycks and Steven Basart and Saurav Kadavath and Mantas Mazeika and Akul Arora and Ethan Guo and Collin Burns and Samir Puranik and Horace He and Dawn Song and Jacob Steinhardt},
  journal={NeurIPS},
  year={2021}
}

@misc{vandervleuten2023drbootbootstrappingprogram,
      title={Dr. Boot: Bootstrapping Program Synthesis Language Models to Perform Repairing}, 
      author={Noah van der Vleuten},
      year={2023},
      eprint={2507.15889},
      archivePrefix={arXiv},
      primaryClass={cs.SE},
      url={https://arxiv.org/abs/2507.15889}, 
}

@inproceedings{munozMeasuringDigitalTwins2024,
  title = {Towards {{Measuring Digital Twins Fidelity}} at {{Runtime}}},
  booktitle = {Proceedings of the {{ACM}}/{{IEEE}} 27th {{International Conference}} on {{Model Driven Engineering Languages}} and {{Systems}}},
  author = {Mu{\~n}oz, Paula and Troya, Javier and Vallecillo, Antonio},
  year = {2024},
  month = oct,
  series = {{{MODELS Companion}} '24},
  pages = {507--512},
  publisher = {Association for Computing Machinery},
  address = {New York, NY, USA},
  doi = {10.1145/3652620.3688267},
  urldate = {2025-10-10},
  isbn = {979-8-4007-0622-6},
}

@article{weerdt2012processalignment,
  title = {Process Diagnostics Using Trace Alignment: {{Opportunities}}, Issues, and Challenges},
  shorttitle = {Process Diagnostics Using Trace Alignment},
  author = {Jagadeesh Chandra Bose, R. P. and {van der Aalst}, Wil M. P.},
  year = {2012},
  month = apr,
  journal = {Information Systems},
  series = {Management and {{Engineering}} of {{Process-Aware Information Systems}}},
  volume = {37},
  number = {2},
  pages = {117--141},
  issn = {0306-4379},
  doi = {10.1016/j.is.2011.08.003},
  urldate = {2025-10-10}
}

@inproceedings{yang2017pima,
  title = {Process-{{Oriented Iterative Multiple Alignment}} for {{Medical Process Mining}}},
  booktitle = {2017 {{IEEE International Conference}} on {{Data Mining Workshops}} ({{ICDMW}})},
  author = {Chen, Shuhong and Yang, Sen and Zhou, Moliang and Burd, Randall and Marsic, Ivan},
  year = {2017},
  month = nov,
  pages = {438--445},
  issn = {2375-9259},
  doi = {10.1109/ICDMW.2017.63},
  urldate = {2025-10-10}
}

@article{es-haghiMethodsEnablingRealtime2024,
  title = {Methods for Enabling Real-Time Analysis in Digital Twins: {{A}} Literature Review},
  shorttitle = {Methods for Enabling Real-Time Analysis in Digital Twins},
  author = {{Es-haghi}, Mohammad Sadegh and Anitescu, Cosmin and Rabczuk, Timon},
  year = {2024},
  month = jul,
  journal = {Computers \& Structures},
  volume = {297},
  pages = {107342},
  issn = {0045-7949},
  doi = {10.1016/j.compstruc.2024.107342},
  urldate = {2025-10-11},
}

@article{iranshahiDigitalTwinsRecent2025,
  title = {Digital Twins: {{Recent}} Advances and Future Directions in Engineering Fields},
  shorttitle = {Digital Twins},
  author = {Iranshahi, Kamran and Brun, Joshua and Arnold, Tim and Sergi, Thomas and M{\"u}ller, Ulf Christian},
  year = {2025},
  month = jun,
  journal = {Intelligent Systems with Applications},
  volume = {26},
  pages = {200516},
  issn = {2667-3053},
  doi = {10.1016/j.iswa.2025.200516},
  urldate = {2025-10-11},
}

@inproceedings{munoz2022using,
  title={Using trace alignments for measuring the similarity between a physical and its digital twin},
  author={Mu{\~n}oz, Paula and Wimmer, Manuel and Troya, Javier and Vallecillo, Antonio},
  booktitle={Proceedings of the 25th international conference on model driven engineering languages and systems: Companion proceedings},
  pages={503--510},
  year={2022}
}

@article{liu2023agentbench,
  title={Agentbench: Evaluating llms as agents},
  author={Liu, Xiao and Yu, Hao and Zhang, Hanchen and Xu, Yifan and Lei, Xuanyu and Lai, Hanyu and Gu, Yu and Ding, Hangliang and Men, Kaiwen and Yang, Kejuan and others},
  journal={arXiv preprint arXiv:2308.03688},
  year={2023}
}

@article{jimenez2023swe,
  title={Swe-bench: Can language models resolve real-world github issues?},
  author={Jimenez, Carlos E and Yang, John and Wettig, Alexander and Yao, Shunyu and Pei, Kexin and Press, Ofir and Narasimhan, Karthik},
  journal={arXiv preprint arXiv:2310.06770},
  year={2023}
}

@inproceedings{eghbali2022crystalbleu,
  title={CrystalBLEU: precisely and efficiently measuring the similarity of code},
  author={Eghbali, Aryaz and Pradel, Michael},
  booktitle={Proceedings of the 37th IEEE/ACM International Conference on Automated Software Engineering},
  pages={1--12},
  year={2022}
}

@article{yang2025leveraging,
  title={Leveraging Large Language Models for Enhanced Digital Twin Modeling: Trends, Methods, and Challenges},
  author={Yang, Linyao and Luo, Shi and Cheng, Xi and Yu, Lei},
  journal={arXiv preprint arXiv:2503.02167},
  year={2025}
}

@article{ershenko2025quantitative,
  title={Quantitative metrics for validation and decision-making in digital twins: a comparative study on a railway braking system},
  author={Ershenko, Dmitrii and Derbysheva, Glafira and Panayi, Andreas and Fortin, Clement},
  journal={Proceedings of the Design Society},
  volume={5},
  pages={2671--2680},
  year={2025},
  publisher={Cambridge University Press}
}

@inproceedings{zhang2024large,
  title={Large language models for explainable decisions in dynamic digital twins},
  author={Zhang, Nan and Vergara-Marcillo, Christian and Diamantopoulos, Georgios and Shen, Jingran and Tziritas, Nikos and Bahsoon, Rami and Theodoropoulos, Georgios},
  booktitle={International Conference on Dynamic Data Driven Applications Systems},
  pages={81--89},
  year={2024},
  organization={Springer}
}

@article{wang2025llm,
  title={An LLM-guided SD-LDM Digital Twin Construction Strategy (LSDT) for multi-industrial scenarios: Enhancing adaptability and efficiency},
  author={Wang, Feixiang and Liu, Xiaojun and Lv, Feng and Wang, Chongxin and Shi, Jin and Zheng, Xiaotian and Li, Chao},
  journal={Journal of Manufacturing Systems},
  volume={80},
  pages={995--1012},
  year={2025},
  publisher={Elsevier}
}

@article{lin2024ddd,
  title={Ddd-gendt: Dynamic data-driven generative digital twin framework},
  author={Lin, Yu-Zheng and Shi, Qinxuan and Yang, Zhanglong and Latibari, Banafsheh Saber and Satam, Shalaka and Shao, Sicong and Salehi, Soheil and Satam, Pratik},
  journal={arXiv preprint arXiv:2501.00051},
  year={2024}
}

\beginsupplement
\section{EnvTrace Framework with Beamline Control-logic Digital Twin}

\begin{figure}[H]
  \centering
  \includegraphics[width=\textwidth, trim = 0cm 3.5cm 0cm 5.0cm]{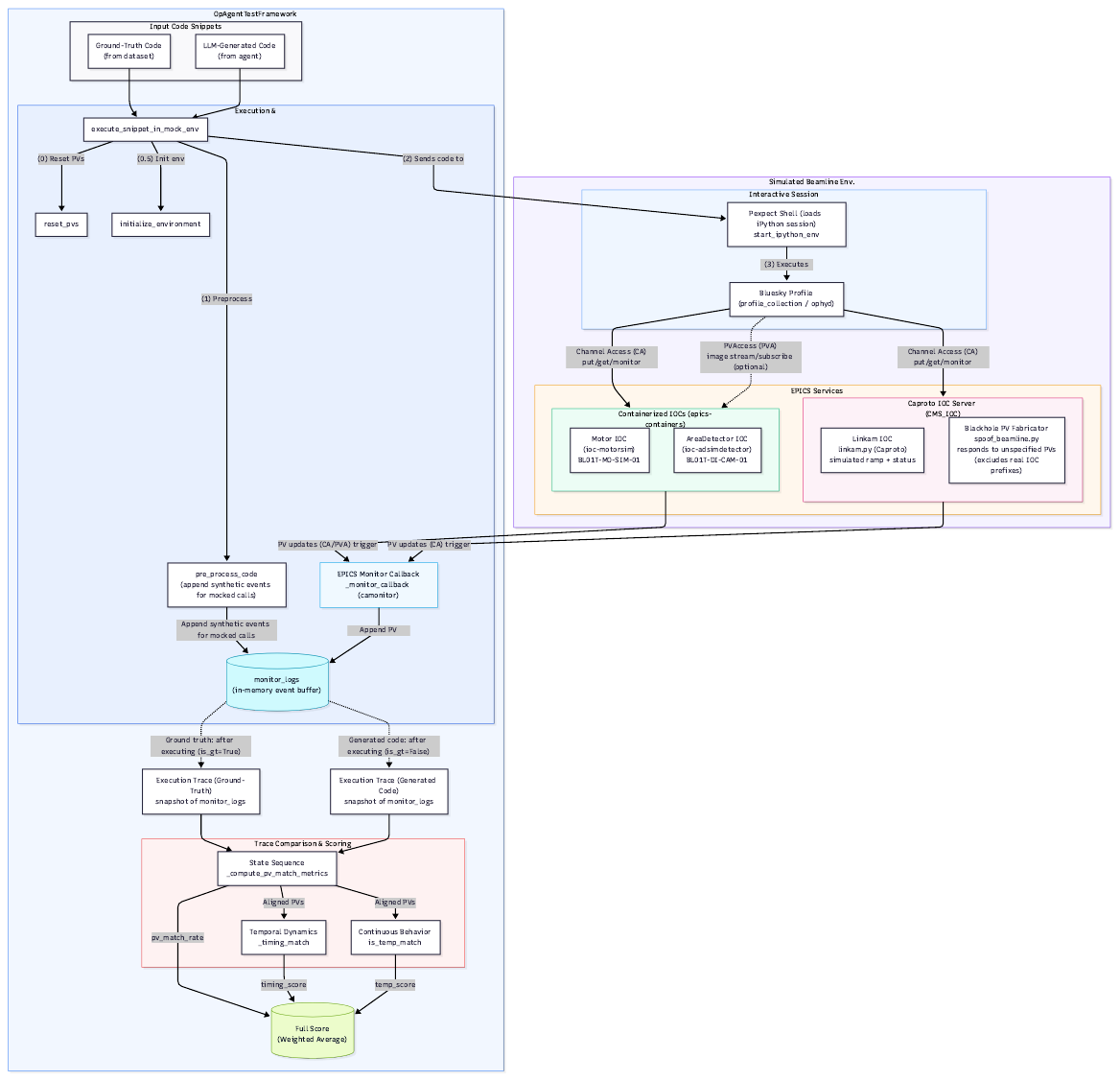}
  \caption{A diagram of the evaluation architecture showing how LLM-generated and ground-truth code run in an IPython session connected to a simulated beamline. PV updates are recorded as execution traces, based on which the EnvTrace framework calculates a comprehensive score reflecting state, timing, and behavioral alignment.}
  \label{si:overview}
\end{figure}

A detailed diagram of the EnvTrace architecture is provided in Fig.~\ref{si:overview}. LLM-generated code and ground-truth code are executed within a sandboxed, interactive IPython session. 
The session, managed via the \texttt{pexpect} library for programmatic CLI interaction, communicates with a simulated beamline environment running EPICS IOCs in Docker containers.
The environment is configured with the same set of Python functions used during live experiments at the beamline, providing access to the exact same Python API for instrument control.
In other words, the same Python scripts used for beamline experiments can be run in the simulation without any modifications.
An EPICS monitor captures all state changes (PV updates), generating an execution trace. The EnvTrace framework then aligns the ground-truth (reference) and predicted traces to compute a multi-faceted EnvTrace accuracy and full score based on state, temporal, and behavioral equivalence.

\section{EnvTrace Criteria}
\label{si:details}
In this work, we monitor the PVs associated with the sample stages, a Linkam thermal stage, and an X-ray detector, which are all major components used at the beamline. Depending on the task and the physical instrument's status or the outcomes, different components can be tracked in EnvTrace to provide the full score and accuracy. Each metric's configuration can be tailored to fit specific use cases.


\subsection{EnvTrace Full Score}
\label{si:continuous_score_formulas}

EnvTrace calculates a continuous, granular \texttt{full\_score} between 0.0 and 1.0. This score provides a more nuanced measure of similarity than a strict pass/fail metric. It is composed of three component scores: \texttt{pv\_match\_rate}, \texttt{timing\_score}, and \texttt{temp\_score}.

\textbf{State Sequence Score (\texttt{pv\_match\_rate})}
The \texttt{pv\_match\_rate}, denoted as $S_{\text{pv}}$, is the fraction of correctly matched state changes after sequence alignment. It is calculated as:
\begin{equation}
S_{\text{pv}} = \frac{N_{\text{value\_matches}}}{N_{\text{total\_pairs}}}
\end{equation}
where $N_{\text{value\_matches}}$ is the number of aligned pairs with identical PV names and values, and $N_{\text{total\_pairs}}$ is the total length of the aligned sequence (including insertions and deletions).

\textbf{Temporal Dynamics Score (\texttt{timing\_score})}
The \texttt{timing\_score}, denoted as $S_{\text{timing}}$, is a weighted average of four sub-scores that quantify different aspects of temporal similarity.
\begin{equation}
S_{\text{timing}} = 0.4 \cdot S_{\text{r2}} + 0.2 \cdot S_{\text{slope}} + 0.2 \cdot S_{\text{duration}} + 0.2 \cdot S_{\text{mape}}
\end{equation}
The component scores are defined as:
\begin{itemize}
    \item {R-squared Score ($S_{\text{r2}}$):} Directly uses the coefficient of determination from the linear regression of timestamps.
    \begin{equation}
    S_{\text{r2}} = R^2
    \end{equation}
    \item {Slope Score ($S_{\text{slope}}$):} A linear penalty for deviations from a perfect slope of 1.0.
    \begin{equation}
    S_{\text{slope}} = \max(0, 1 - |\text{slope} - 1|)
    \end{equation}
    \item {Duration Score ($S_{\text{duration}}$):} A linear penalty based on the relative difference in total duration, normalized by a tolerance $\tau_{\text{dur}} = 0.25$.
    \begin{equation}
    S_{\text{duration}} = \max\left(0, 1 - \frac{|\text{dur}_{\text{pred}} - \text{dur}_{\text{gt}}| / \text{dur}_{\text{gt}}}{\tau_{\text{dur}}}\right)
    \end{equation}
    \item {MAPE Score ($S_{\text{mape}}$):} A linear penalty based on the Mean Absolute Percentage Error of the intervals, normalized by a tolerance $\tau_{\text{mape}} = 1.0$.
    \begin{equation}
    S_{\text{mape}} = \max\left(0, 1 - \frac{\text{MAPE}}{\tau_{\text{mape}}}\right)
    \end{equation}
\end{itemize}

\textbf{Continuous Process Score (\texttt{temp\_score})}
The \texttt{temp\_score}, denoted as $S_{\text{temp}}$, is a weighted average of two sub-scores based on an exponential decay function. This function penalizes large errors more significantly than small ones, providing a smooth score between 0 and 1.
\begin{equation}
S_{\text{temp}} = 0.7 \cdot S_{\text{mae}} + 0.3 \cdot S_{\text{final\_temp}}
\end{equation}
The component scores are defined as:
\begin{itemize}
    \item {MAE Score ($S_{\text{mae}}$):} An exponential decay score based on the Mean Absolute Error of the temperature profiles.
    \begin{equation}
    S_{\text{mae}} = \exp\left(-\frac{\text{MAE}}{\lambda}\right)
    \end{equation}
    \item {Final Temperature Score ($S_{\text{final\_temp}}$):} An exponential decay score based on the absolute difference in the final temperatures, $\Delta T_{\text{final}}$.
    \begin{equation}
    S_{\text{final\_temp}} = \exp\left(-\frac{\Delta T_{\text{final}}}{\lambda}\right)
    \end{equation}
\end{itemize}
For both scores, the characteristic scale $\lambda$ is set to \textbf{15.0$^{\circ}$C}.

\textbf{Holistic Full Score}
The final continuous EnvTrace \texttt{full\_score}, denoted as $S_{\text{full}}$, is the weighted sum of the component scores, as described also in Eq.~(\ref{eq:full_score}).
\begin{equation}
S_{\text{full}} = 
\begin{cases} 
    0.6 \cdot S_{\text{pv}} + 0.2 \cdot S_{\text{timing}} + 0.2 \cdot S_{\text{temp}} & \text{if temperature involved} \\
    0.8 \cdot S_{\text{pv}} + 0.2 \cdot S_{\text{timing}} & \text{if no temperature involved}
\end{cases}
\end{equation}


\subsection{EnvTrace Accuracy}
\label{si:accuracy_details}

The binary ``EnvTrace accuracy'' metric is a composite metric derived from three binary flags: \texttt{exact\_pv\_match}, \texttt{timing\_match}, and \texttt{temp\_match}. A code snippet is only considered a functional match and thus high ``accuracy'' if all applicable criteria are met. The logic is as follows:

\begin{equation}
\label{eq:full_match_logic}
\texttt{accuracy} = 
\begin{cases}
    \texttt{exact\_pv\_match} \land \texttt{timing\_match} \land \texttt{temp\_match} & \text{if temperature involved} \\
    \texttt{exact\_pv\_match} \land \texttt{timing\_match} & \text{if no temperature involved}
\end{cases}
\end{equation}

The specific criteria for each of these binary flags are detailed below.

\textbf{State Sequence Match (\texttt{exact\_pv\_match})}
The \texttt{exact\_pv\_match} criterion evaluates whether the sequence of non-temperature-related state changes is perfectly identical between the ground-truth and predicted execution traces. It is the most stringent component of the evaluation. This flag is set to \texttt{True} if and only if:
\begin{itemize}
    \item The number of PV changes in the predicted trace is exactly equal to the number of PV changes in the ground-truth trace.
    \item After sequence alignment, there are no insertions or deletions (i.e., no extra or missing operations).
    \item For every aligned pair of PV changes, the PV name and its corresponding value are identical (within a floating-point tolerance of $1 \times 10^{-3}$ for numerical values).
\end{itemize}
This ensures a perfect one-to-one correspondence in both the actions performed and their outcomes.

\textbf{Temporal Dynamics Match (\texttt{timing\_match})}
The \texttt{timing\_match} criterion assesses whether the execution was paced correctly. This is determined by performing a linear regression on the timestamps of the correctly matched PV events. The flag is set to \texttt{True} if all of the following conditions are met:
\begin{itemize}
    \item {Goodness of Fit:} The coefficient of determination ($R^2$) from the linear regression must be greater than or equal to a threshold of \textbf{0.90}. This ensures a consistent, linear relationship between the timing of the two executions.
    \item {Pacing Slope:} The slope of the regression line must be within the range [\textbf{0.8}, \textbf{1.2}]. A slope of 1.0 indicates identical pacing; this range allows for a minor (20\%) deviation.
    \item {Total Duration:} The relative difference in the total duration of the event sequence must be less than or equal to a tolerance of \textbf{25\%}.
    \item {Interval Consistency:} The Mean Absolute Percentage Error (MAPE) between the time intervals of consecutive events must be less than or equal to a tolerance of \textbf{100\%}.
\end{itemize}

\textbf{Continuous Process Match (\texttt{temp\_match})}
For experiments involving temperature ramps, the \texttt{temp\_match} criterion evaluates the fidelity of the temperature profile. The flag is set to \texttt{True} if both of the following conditions are met:
\begin{itemize}
    \item {Mean Absolute Error (MAE):} The MAE between the ground-truth and predicted temperature logs must be less than or equal to a threshold of \textbf{5.0$^{\circ}$C}.
    \item {Final Temperature Difference:} The absolute difference between the final recorded temperatures in the two logs must be less than or equal to a threshold of \textbf{5.0$^{\circ}$C}.
\end{itemize}

\newpage

\section{Models: Date, Provider, Quantization, Temperature}
\label{si:models}

For assessing code quality and benchmarking model performance, models from a range of closed-source vendors and open-source communities were included, ensuring that the results encompassed both proprietary systems and publicly available alternatives.
A sampling temperature of 0 was used for most models to promote deterministic outputs.  
For models with 'thinking' or reasoning capabilities explicitly enabled (e.g., \texttt{Claude-Sonnet-4-(Thinking)}), a temperature of 1.0 was required. 

\begin{threeparttable}
\label{si:model_info}
\caption{Models: Date, Provider, Quantization, Temperature}
\begin{tabular}{lcccc}
\toprule
Model & Date & Provider & Temperature & Reasoning Token Budget \\
\midrule
\multicolumn{5}{l}{\textbf{Closed Source Models}} \\
\midrule
\multicolumn{5}{l}{\textit{Anthropic}} \\
  Claude Sonnet 4 (Thinking) & 2025-05-14 & Native API & 1 & 2000 \\
  Claude Opus 4 (Thinking) & 2025-05-14 & Native API & 1 & 2000 \\
  Claude 3.5 Sonnet & 2024-10-22 & Native API & 0 & \multicolumn{1}{c}{—} \\
  Claude Opus 4 (Abacus) & (2025-08-06) & Abacus & 0 & \multicolumn{1}{c}{—} \\
  Claude Sonnet 4 (Abacus) & (2025-08-06) & Abacus & 0 & \multicolumn{1}{c}{—} \\
  Claude Sonnet 4 (Bedrock) & 2025-05-14 & Bedrock & 0 & \multicolumn{1}{c}{—} \\
\midrule
\multicolumn{5}{l}{\textit{Google}} \\
  Gemini 2.5 Pro & (2025-09-15) & Native API & 0 &  8192 \\
\midrule
\multicolumn{5}{l}{\textit{xAI}} \\
  Grok-4 Fast & (2025-09-22) & OpenRouter & 0 & \multicolumn{1}{c}{—} \\
  Grok-4 & (2025-09-16) & OpenRouter & 0 & \multicolumn{1}{c}{—} \\
  Grok Code Fast 1 & (2025-09-17) & OpenRouter & 0 & \multicolumn{1}{c}{—} \\
\midrule
\multicolumn{5}{l}{\textit{OpenAI / Microsoft}} \\
  GPT-5 (high) & 2025-08-07 & Azure & 1 & high\\
  GPT-4o & 2024-05-13 & Azure & 0 & \multicolumn{1}{c}{—} \\
  o3 (high) & 2025-04-16 & Azure & 1 & high \\
  GPT-5 (minimal) & 2025-08-07 & Azure & 1 & minimal \\
  GPT-4o (Abacus) & (2025-08-01) & Abacus & 0 & \multicolumn{1}{c}{—} \\
\midrule
\multicolumn{4}{l}{\textit{Mistral}} \\
  Devstral Medium & 2024-07-10 & OpenRouter & 0 & \multicolumn{1}{c}{—} \\
\bottomrule
\end{tabular}
\begin{tablenotes}[flushleft]
\footnotesize
\item Date in parentheses indicates access date due to unknown model date.
\end{tablenotes}
\end{threeparttable}

\bigskip

\begin{threeparttable}
\begin{tabular}{lccc}
\toprule
Model & Quantization & Provider & Temperature \\
\midrule
\multicolumn{4}{l}{\textbf{Open Source Models}} \\
\midrule
  Qwen3-Coder (480B) & fp8/fp4 & OpenRouter & 0 \\
  Athene-v2 (72.7B) & Q4\_K\_M & Ollama & 0 \\
  Qwen3-Coder (30.5B) & Q4\_K\_M & Ollama & 0 \\
  Athene-v2-Agent (72.7B) & Q4\_K\_M & Hugging Face\tnote{a} & 0 \\
  Llama3.3 (70.6B) & Q4\_K\_M & Ollama & 0 \\
  Qwen2.5 (7.62B) & Q4\_K\_M & Ollama & 0 \\
  Qwen2.5-Coder (32.8B) & Q4\_K\_M & Ollama & 0 \\
  Mistral-NeMo (12.2B) & Q4\_0 & Ollama & 0 \\
  GPT-oss (117B) & MXFP4 & Ollama & 0 \\
  Qwen2 (7.62B) & Q4\_0 & Ollama & 0 \\
  Mistral (7.3B) & Q4\_0 & Ollama & 0 \\
  Qwen3 (32B) & Q4\_K\_M & Ollama & 0 \\
  Phi-3.5 (3.8B, fp16) & fp16 & Ollama & 0 \\
  Phi-3.5 (3.8B) & Q4\_0 & Ollama & 0 \\
\bottomrule
\end{tabular}
\begin{tablenotes}[flushleft]
\footnotesize
\item[a] Available at: \tiny\url{https://huggingface.co/lmstudio-community/Athene-V2-Agent-GGUF}
\end{tablenotes}
\end{threeparttable}

\section{Short system prompt}
\label{si:baseline_performance}

To evaluate how prompt length influences model performance, we performed the EnvTrace analysis using a short system prompt that is similar to the one used in \citet{mathur2024vision}. 
In contrast to the 5,000-word prompt ($\sim$8500 tokens with \texttt{GPT-4o} tokenizer) with extensive markdown formatting used in Section~\ref{sec:results}, this shorter version contains about 2,000 words ($\sim$3,800 tokens) and consists of only the essential but not all routine beamline functions.
Table~\ref{tab:simple_results_old_prompt_new_data} summarizes the results for the 116 simple-flow tasks using this short prompt.
The open-source models perform substantially better under this short prompt than under the longer prompt used Table~\ref{tab:simple_results} in Section~\ref{sec:results}.  
Simpler, less token-heavy instructions reduce context overhead, which appears to benefit smaller models such as \texttt{Qwen2.5-Coder}, \texttt{Llama3.3}, and \texttt{Athene-v2}.  
With the short prompt, these models rival the closed-source models \texttt{Claude-3.5-Sonnet} and \texttt{GPT-4o}.  
Table~\ref{tab:complex_results_old_prompt} shows the results for the 20 complex-flow code generation tasks with the short system prompt, for comparison with the results from the long prompt in Table~\ref{tab:complex_results}.  
Top-performing reasoning-enabled models, such as \texttt{Claude-Sonnet-4-(Thinking)} and \texttt{GPT-4o}, achieve high full scores (above 85\%) and functional accuracies comparable to human benchmarks, while open-source models show mixed performance.  
Although smaller models remain weaker in logical control and looping structures, with the short prompt they produced syntactically valid and partially functional code much more consistently than with the longer prompt.
Overall, these results indicate that simpler, shorter prompts enhance the reliability of open-source LLMs in instrument control tasks.  

\begin{table}[htbp]
  \centering
  \begin{threeparttable}
    \caption{Performance of LLMs on the new dataset (N=116) using the short system prompt.}
    \label{tab:simple_results_old_prompt_new_data}
    \sisetup{separate-uncertainty=true, table-align-uncertainty=true}
    \begin{tabular}{
      l
      S[table-format=2.2(2.2)]
      S[table-format=2.2(2.2)]
      S[table-format=2.2(2.2)]
      S[table-format=2.2(2.2)]
      S[table-format=2.2(2.2)]
    }
      \toprule
      \textbf{Model} & 
      {\textbf{\begin{tabular}[c]{@{}c@{}}EnvTrace \\ Full Score (\%, $\uparrow$)\tnote{a}\end{tabular}}} & 

      {\textbf{\begin{tabular}[c]{@{}c@{}} EnvTrace \\ Accuracy (\%, $\uparrow$)\tnote{b}\end{tabular}}} &
      {\textbf{\begin{tabular}[c]{@{}c@{}} \textcolor{gray}{Exact Match} \\  \textcolor{gray}{(\%, $\uparrow$)\tnote{c}}\end{tabular}}} & 
      {\textbf{\begin{tabular}[d]{@{}c@{}}\textcolor{gray}{Norm. Lev.} \\ \textcolor{gray}{Dist.} \textcolor{gray}{(\%, $\downarrow$)\tnote{d}}\end{tabular}}} &
      {\textbf{\begin{tabular}[c]{@{}c@{}}Inference \\ Time (s, $\downarrow$)\end{tabular}}} \\
      \midrule
\multicolumn{6}{l}{\textbf{Open Source Models}} \\
\midrule
  Athene-v2-Agent (72.7B) & 100.0 \pm 0.0 & 100.0 \pm 0.0 & 95.7 \pm 0.0 & 0.8 \pm 0.0 & 0.7 \pm 0.1 \\
  Athene-v2 (72.7B) & 99.7 \pm 0.0 & 99.1 \pm 0.0 & 94.8 \pm 0.0 & 1.0 \pm 0.0 & 0.8 \pm 0.1 \\
  Llama3.3 (70.6B) & 99.5 \pm 0.1 & 98.6 \pm 0.5 & 89.1 \pm 0.5 & 2.5 \pm 0.2 & 0.9 \pm 0.1 \\
  Qwen2.5-Coder (32.8B) & 98.5 \pm 0.0 & 95.7 \pm 0.0 & 89.7 \pm 0.0 & 2.8 \pm 0.0 & 1.0 \pm 0.0 \\
  Mistral-NeMo (12.2B) & 97.5 \pm 0.0 & 92.2 \pm 0.0 & 77.3 \pm 1.0 & 5.6 \pm 0.2 & 1.2 \pm 0.1 \\
  Qwen2.5 (7.62B) & 96.0 \pm 0.0 & 94.0 \pm 0.0 & 85.9 \pm 0.5 & 5.1 \pm 0.1 & 1.1 \pm 0.1 \\
  Qwen2 (7.62B) & 95.2 \pm 0.0 & 91.4 \pm 0.0 & 81.0 \pm 0.0 & 6.2 \pm 0.0 & 1.1 \pm 0.0 \\
  Mistral (7.3B) & 41.3 \pm 0.0 & 37.9 \pm 0.0 & 19.0 \pm 0.0 & 46.3 \pm 0.0 & 1.3 \pm 0.0 \\
  Phi-3.5 (3.8B, fp16) & 14.0 \pm 2.9 & 0.9 \pm 1.5 & 0.9 \pm 1.5 & 98.8 \pm 2.0 & 1.2 \pm 1.1 \\
  Phi-3.5 (3.8B) & 13.6 \pm 3.5 & 0.9 \pm 1.5 & 0.9 \pm 1.5 & 98.4 \pm 2.7 & 0.8 \pm 0.4 \\
\midrule
\multicolumn{6}{l}{\textbf{Closed Source Models}} \\
\midrule
\multicolumn{6}{l}{\textit{Anthropic}} \\
  Claude 3.5 Sonnet & 99.5 \pm 0.0 & 98.3 \pm 0.0 & 93.1 \pm 0.0 & 1.4 \pm 0.0 & 0.9 \pm 0.0 \\
  Claude Sonnet 4 (Thinking)\tnote{t} & 97.5 \pm 0.4 & 96.3 \pm 0.5 & 92.5 \pm 1.3 & 2.7 \pm 0.3 & 4.5 \pm 0.3 \\
\midrule
\multicolumn{6}{l}{\textit{OpenAI / Microsoft}} \\
  GPT-4o & 99.4 \pm 0.1 & 96.0 \pm 0.5 & 89.9 \pm 0.5 & 2.2 \pm 0.2 & 0.7 \pm 0.1 \\
\bottomrule
    \end{tabular}
    \begin{tablenotes}
      \small
      \item[a] \textbf{EnvTrace Full Score (\%):} The continuous score from 0 to 100, reflecting the weighted average of state and temporal fidelity.
      \item[b] \textbf{EnvTrace Accuracy (\%):} The percentage of test cases passing the strict binary criteria for semantic equivalence.\item[c] \textbf{Exact Match (\%):} The percentage of test cases where the generated code string is identical to a ground-truth solution \cite{mathur2024vision}.
        \item[d] \textbf{Normalized Levenshtein Distance (\%):} A value of $0\%$ indicates identical strings.
      \item[e] All results are the mean and standard deviation over three runs. \item[f] Models are grouped and then sorted by descending Average Full Score.
    \end{tablenotes}
  \end{threeparttable}
\end{table}

\begin{table}[htbp]
  \centering
  \begin{threeparttable}
    \caption{Performance of LLMs on Complex-Flow Code Generation Tasks (N=20) with the short prompt}
    \label{tab:complex_results_old_prompt}
    \sisetup{separate-uncertainty=true, table-align-uncertainty=true}
    \begin{tabular}{
      l
      S[table-format=2.2(2.2)]
      S[table-format=2.2(2.2)]
      S[table-format=2.2(2.2)]
      S[table-format=3.2(2.2)]
      S[table-format=3.2(2.2)]
    }
      \toprule
      \textbf{Model} & 
      {\textbf{\begin{tabular}[c]{@{}c@{}}EnvTrace \\ Full Score (\%, $\uparrow$)\tnote{a}\end{tabular}}} & 
      {\textbf{\begin{tabular}[c]{@{}c@{}}EnvTrace \\ Accuracy (\%, $\uparrow$)\tnote{b}\end{tabular}}} & 
      \textcolor{gray}{\textbf{\begin{tabular}[c]{@{}c@{}}CodeBLEU \\ (comb-7, \%, $\uparrow$)\tnote{c}\end{tabular}}} &
      \textcolor{gray}{\textbf{\begin{tabular}[c]{@{}c@{}}Norm. Lev. \\ Dist. (\%, $\downarrow$)\tnote{d}\end{tabular}}} &
      {\textbf{\begin{tabular}[c]{@{}c@{}}Inference \\ Time (s)\end{tabular}}} \\
\midrule
\multicolumn{6}{l}{\textbf{Closed Source Models}} \\
\midrule
\multicolumn{6}{l}{\textit{Anthropic}} \\
  Claude Sonnet 4 (Thinking)\tnote{t} & 91.9 \pm 1.5 & 55.0 \pm 5.0 & 55.5 \pm 1.1 & 42.0 \pm 3.3 & 9.4 \pm 0.7 \\
  Claude 3.5 Sonnet & 89.8 \pm 0.5 & 61.7 \pm 2.9 & 52.4 \pm 0.1 & 39.4 \pm 0.2 & 2.1 \pm 0.1 \\
\midrule
\multicolumn{6}{l}{\textit{OpenAI / Microsoft}} \\
  GPT-4o & 83.0 \pm 0.9 & 43.3 \pm 2.9 & 52.2 \pm 0.2 & 44.9 \pm 0.9 & 1.5 \pm 0.0 \\
\midrule
\multicolumn{6}{l}{\textbf{Open Source Models}} \\
\midrule
  Athene-v2-Agent (72.7B) & 79.5 \pm 0.0 & 40.0 \pm 0.0 & 48.2 \pm 0.0 & 46.1 \pm 0.0 & 7.0 \pm 0.5 \\
  Athene-v2 (72.7B) & 72.2 \pm 0.2 & 41.7 \pm 2.9 & 53.0 \pm 0.0 & 45.3 \pm 0.2 & 7.2 \pm 0.1 \\
  Qwen2.5-Coder (32.8B) & 72.0 \pm 0.0 & 45.0 \pm 0.0 & 51.3 \pm 0.0 & 44.7 \pm 0.0 & 3.9 \pm 0.4 \\
  Llama3.3 (70.6B) & 71.3 \pm 0.1 & 25.0 \pm 0.0 & 50.0 \pm 0.0 & 49.3 \pm 0.1 & 6.1 \pm 0.5 \\
  Qwen2 (7.62B) & 62.4 \pm 0.9 & 25.0 \pm 0.0 & 47.1 \pm 1.1 & 50.7 \pm 0.5 & 1.2 \pm 0.3 \\
  Mistral-NeMo (12.2B) & 58.0 \pm 2.4 & 26.7 \pm 5.8 & 43.9 \pm 1.4 & 48.3 \pm 2.5 & 1.8 \pm 0.3 \\
  Qwen2.5 (7.62B) & 52.1 \pm 1.0 & 10.0 \pm 0.0 & 48.2 \pm 0.3 & 50.6 \pm 0.4 & 1.7 \pm 0.5 \\
  Mistral (7.3B) & 2.8 \pm 0.2 & 0.0 \pm 0.0 & 48.2 \pm 0.0 & 66.8 \pm 0.0 & 1.8 \pm 0.0 \\
  Phi-3.5 (3.8B) & 2.0 \pm 0.0 & 0.0 \pm 0.0 & 40.0 \pm 0.0 & 100.0 \pm 0.0 & 0.6 \pm 0.0 \\
  Phi-3.5 (3.8B, fp16) & 2.0 \pm 0.0 & 0.0 \pm 0.0 & 40.0 \pm 0.0 & 100.0 \pm 0.0 & 0.5 \pm 0.0 \\
\bottomrule
    \end{tabular}
    \begin{tablenotes}
      \small
      \item[a] \textbf{EnvTrace Full Score (\%):} The continuous score from 0 to 100, reflecting the weighted average of state and temporal fidelity.
      \item[b] \textbf{EnvTrace Accuracy (\%):} The percentage of test cases passing the strict binary criteria for semantic equivalence.
      \item[c] \textbf{CodeBLEU (comb-7, \%):} syntactic similarity metric. We report the ``comb-7'' variant, which weights data-flow and syntax more heavily \cite{ren2020codebleu}.
      \item[d] \textbf{Normalized Levenshtein Distance (\%):} A value of $0\%$ indicates identical strings.
      \item[e] All results are the mean and standard deviation over three runs. \item[f] Models are grouped and then sorted by descending Average Full Score.
      \item[h] \textbf{Human Inference Time} in the order of minutes (>100s)
      \item[t] Models run with reasoning enabled. ``(high)'' and ``(minimal)'' denote different reasoning levels.
    \end{tablenotes}
  \end{threeparttable}
\end{table}


\section{Complex-flow results}
\label{si:complex}
An example of PV trace alignment is shown to demonstrate the calculation of the PV match rate.
We also present a detailed breakdown of LLM performance by PV match, timing, and temperature, offering deeper insight into the model's coding capabilities.

\subsection{PV match example}
\label{si:pvmatch}
For the \texttt{GPT-5-high} example on map scan in Box~\ref{lst:complex}, a comparison on the PV traces is given in Box~\ref{lst:PVcomplex}. The 'check' and 'cross' in the right-most column show whether the PV traces match or not. Although both code versions are semantically correct and perform map scans, differences in the scanning order led to mismatched PV traces. This case illustrates a situation where the scoring reflects a mismatch rather than a true error.

\begin{lstlisting}[
float,
  basicstyle=\ttfamily\fontsize{3.5}{6}\selectfont,
  columns=fullflexible, keepspaces=true, breaklines=false
keepspaces=true,
  tabsize=4,
  showstringspaces=false,
  upquote=true,
  literate={✓}{{\cmarkbox}}1 {✗}{{\xmarkbox}}1
           {–}{{-}}1 {—}{{-}}1 {…}{{...}}3,
 caption={Trace alignment and scoring for the \texttt{GPT-5-high} map scan example shown in Box~\ref{lst:complex}.},
 label={lst:PVcomplex}
]
COMMAND:
--------------------------------------------------------------------------------
Do a map scan, x range from 0 to 0.3mm, y from 0 to 0.6mm, step size is 0.15 horizontally and 0.2 vertically. (Exposure time 1s.)
--------------------------------------------------------------------------------

====================================================================================================
                                   LOGS COMPARISON
====================================================================================================
-----------------------------------------------------------------------------------------------------------------------------------------------------------------------------
                                    GROUND TRUTH                                      |                                       PREDICTED
-----------------------------------------------------------------------------------------------------------------------------------------------------------------------------
| XF:11BMB-ES{Det:PIL2M}:cam1:AcquireTime | 09:24:33.402     | 1.0                   | | | XF:11BMB-ES{Det:PIL2M}:cam1:AcquireTime | 18:07:23.020     | 1.0                   | ✓
| XF:11BMB-ES{Det:PIL2M}:cam1:Acquire     | 09:24:36.417     | 1                     | | | XF:11BMB-ES{Det:PIL2M}:cam1:Acquire     | 18:07:23.782     | 1                     | ✓
| XF:11BMB-ES{Det:PIL2M}:cam1:Acquire     | 09:24:37.419     | 0                     | | | XF:11BMB-ES{Det:PIL2M}:cam1:Acquire     | 18:07:24.784     | 0                     | ✓
| XF:11BMB-ES{Chm:Smpl-Ax:Z}Mtr           | 09:24:38.189     | 0.2                   | | | XF:11BMB-ES{Chm:Smpl-Ax:X}Mtr           | 18:07:25.335     | 0.15                  | ✗
| XF:11BMB-ES{Det:PIL2M}:cam1:Acquire     | 09:24:44.094     | 1                     | | | XF:11BMB-ES{Det:PIL2M}:cam1:Acquire     | 18:07:30.777     | 1                     | ✓
| XF:11BMB-ES{Det:PIL2M}:cam1:Acquire     | 09:24:45.096     | 0                     | | | XF:11BMB-ES{Det:PIL2M}:cam1:Acquire     | 18:07:31.780     | 0                     | ✓
| XF:11BMB-ES{Chm:Smpl-Ax:Z}Mtr           | 09:24:45.735     | 0.4                   | | | XF:11BMB-ES{Chm:Smpl-Ax:X}Mtr           | 18:07:32.326     | 0.3000000000000004    | ✗
| XF:11BMB-ES{Det:PIL2M}:cam1:Acquire     | 09:24:51.604     | 1                     | | | XF:11BMB-ES{Det:PIL2M}:cam1:Acquire     | 18:07:37.759     | 1                     | ✓
| XF:11BMB-ES{Det:PIL2M}:cam1:Acquire     | 09:24:52.606     | 0                     | | | XF:11BMB-ES{Det:PIL2M}:cam1:Acquire     | 18:07:38.761     | 0                     | ✓
| XF:11BMB-ES{Chm:Smpl-Ax:Z}Mtr           | 09:24:53.235     | 0.6000000000000001    | | | XF:11BMB-ES{Chm:Smpl-Ax:Z}Mtr           | 18:07:39.313     | 0.2                   | ✗
| XF:11BMB-ES{Det:PIL2M}:cam1:Acquire     | 09:24:59.101     | 1                     | | | XF:11BMB-ES{Det:PIL2M}:cam1:Acquire     | 18:07:45.109     | 1                     | ✓
| XF:11BMB-ES{Det:PIL2M}:cam1:Acquire     | 09:25:00.104     | 0                     | | | XF:11BMB-ES{Det:PIL2M}:cam1:Acquire     | 18:07:46.112     | 0                     | ✓
| XF:11BMB-ES{Chm:Smpl-Ax:X}Mtr           | 09:25:00.757     | 0.15                  | | | XF:11BMB-ES{Chm:Smpl-Ax:X}Mtr           | 18:07:46.670     | 0.15000000000000072   | ✓
| XF:11BMB-ES{Chm:Smpl-Ax:Z}Mtr           | 09:25:01.110     | 0.0                   | |                                                                                       ✗
| XF:11BMB-ES{Det:PIL2M}:cam1:Acquire     | 09:25:07.342     | 1                     | | | XF:11BMB-ES{Det:PIL2M}:cam1:Acquire     | 18:07:52.094     | 1                     | ✓
| XF:11BMB-ES{Det:PIL2M}:cam1:Acquire     | 09:25:08.345     | 0                     | | | XF:11BMB-ES{Det:PIL2M}:cam1:Acquire     | 18:07:53.097     | 0                     | ✓
| XF:11BMB-ES{Chm:Smpl-Ax:Z}Mtr           | 09:25:08.980     | 0.2                   | | | XF:11BMB-ES{Chm:Smpl-Ax:X}Mtr           | 18:07:53.646     | 3.608224830031759e-16 | ✗
| XF:11BMB-ES{Det:PIL2M}:cam1:Acquire     | 09:25:14.870     | 1                     | |                                                                                       ✗
| XF:11BMB-ES{Det:PIL2M}:cam1:Acquire     | 09:25:15.873     | 0                     | |                                                                                       ✗
| XF:11BMB-ES{Chm:Smpl-Ax:Z}Mtr           | 09:25:16.502     | 0.4                   | |                                                                                       ✗
| XF:11BMB-ES{Det:PIL2M}:cam1:Acquire     | 09:25:22.361     | 1                     | | | XF:11BMB-ES{Det:PIL2M}:cam1:Acquire     | 18:07:59.070     | 1                     | ✓
| XF:11BMB-ES{Det:PIL2M}:cam1:Acquire     | 09:25:23.364     | 0                     | | | XF:11BMB-ES{Det:PIL2M}:cam1:Acquire     | 18:08:00.072     | 0                     | ✓
| XF:11BMB-ES{Chm:Smpl-Ax:Z}Mtr           | 09:25:23.986     | 0.6000000000000001    | | | XF:11BMB-ES{Chm:Smpl-Ax:Z}Mtr           | 18:08:00.617     | 0.4                   | ✗
| XF:11BMB-ES{Det:PIL2M}:cam1:Acquire     | 09:25:29.852     | 1                     | | | XF:11BMB-ES{Det:PIL2M}:cam1:Acquire     | 18:08:06.349     | 1                     | ✓
| XF:11BMB-ES{Det:PIL2M}:cam1:Acquire     | 09:25:30.855     | 0                     | | | XF:11BMB-ES{Det:PIL2M}:cam1:Acquire     | 18:08:07.352     | 0                     | ✓
| XF:11BMB-ES{Chm:Smpl-Ax:X}Mtr           | 09:25:31.497     | 0.3                   | | | XF:11BMB-ES{Chm:Smpl-Ax:X}Mtr           | 18:08:07.909     | 0.15                  | ✗
| XF:11BMB-ES{Chm:Smpl-Ax:Z}Mtr           | 09:25:32.001     | 0.0                   | | | XF:11BMB-ES{Det:PIL2M}:cam1:Acquire     | 18:08:13.339     | 1                     | ✗
                                                                                      | | XF:11BMB-ES{Det:PIL2M}:cam1:Acquire     | 18:08:14.342     | 0                     | ✗
                                                                                      | | XF:11BMB-ES{Chm:Smpl-Ax:X}Mtr           | 18:08:14.887     | 0.3000000000000004    | ✗
| XF:11BMB-ES{Det:PIL2M}:cam1:Acquire     | 09:25:38.266     | 1                     | | | XF:11BMB-ES{Det:PIL2M}:cam1:Acquire     | 18:08:20.314     | 1                     | ✓
| XF:11BMB-ES{Det:PIL2M}:cam1:Acquire     | 09:25:39.269     | 0                     | | | XF:11BMB-ES{Det:PIL2M}:cam1:Acquire     | 18:08:21.317     | 0                     | ✓
| XF:11BMB-ES{Chm:Smpl-Ax:Z}Mtr           | 09:25:39.915     | 0.2                   | | | XF:11BMB-ES{Chm:Smpl-Ax:Z}Mtr           | 18:08:21.871     | 0.6000000000000001    | ✗
| XF:11BMB-ES{Det:PIL2M}:cam1:Acquire     | 09:25:45.768     | 1                     | | | XF:11BMB-ES{Det:PIL2M}:cam1:Acquire     | 18:08:27.650     | 1                     | ✓
| XF:11BMB-ES{Det:PIL2M}:cam1:Acquire     | 09:25:46.772     | 0                     | | | XF:11BMB-ES{Det:PIL2M}:cam1:Acquire     | 18:08:28.653     | 0                     | ✓
| XF:11BMB-ES{Chm:Smpl-Ax:Z}Mtr           | 09:25:47.401     | 0.4                   | | | XF:11BMB-ES{Chm:Smpl-Ax:X}Mtr           | 18:08:29.199     | 0.15000000000000072   | ✗
| XF:11BMB-ES{Det:PIL2M}:cam1:Acquire     | 09:25:53.206     | 1                     | | | XF:11BMB-ES{Det:PIL2M}:cam1:Acquire     | 18:08:34.625     | 1                     | ✓
| XF:11BMB-ES{Det:PIL2M}:cam1:Acquire     | 09:25:54.209     | 0                     | | | XF:11BMB-ES{Det:PIL2M}:cam1:Acquire     | 18:08:35.628     | 0                     | ✓
| XF:11BMB-ES{Chm:Smpl-Ax:Z}Mtr           | 09:25:54.827     | 0.6000000000000001    | | | XF:11BMB-ES{Chm:Smpl-Ax:X}Mtr           | 18:08:36.175     | 3.608224830031759e-16 | ✗
| XF:11BMB-ES{Det:PIL2M}:cam1:Acquire     | 09:26:00.959     | 1                     | | | XF:11BMB-ES{Det:PIL2M}:cam1:Acquire     | 18:08:41.613     | 1                     | ✓
| XF:11BMB-ES{Det:PIL2M}:cam1:Acquire     | 09:26:01.962     | 0                     | | | XF:11BMB-ES{Det:PIL2M}:cam1:Acquire     | 18:08:42.616     | 0                     | ✓
-----------------------------------------------------------------------------------------------------------------------------------------------------------------------------

SUMMARY:
  Ground Truth: 38 log entries
  Predicted:    36 log entries
  Matches:      24
  Mismatches:   16
  Difference:   2 entries
====================================================================================================

Exact PV match (non-temp): False
PV match rate (non-temp): 60.00%
PV mismatch rate (non-temp): 40.00%
Timing match: True (score: 0.871)
Full match: False (score: 0.654)

\end{lstlisting}

\subsection{Complex-flow results with full score breakdown}
\label{si:breakdown}

The continuous full score and its components provide a deeper diagnostic insight into \textit{why} models succeed or fail. 
Figure~\ref{fig:complex_tasks_components} breaks down the full score for a selection of models on complex flow tasks into its constituent parts: PV Match Rate, Timing Score, and Temperature Score.
The analysis shows that top-performing models like \texttt{Claude} and \texttt{GPT} achieve high, balanced scores across all components, indicating comprehensive success in sequencing, timing, and continuous tracking.
In contrast, for lower-performing models, the primary failure mode is a low PV match rate. For example, \texttt{Llama-3.3} and \texttt{Qwen-2.5} have PV Match scores below 50\%, indicating that they consistently fail to generate the correct sequence of operations. Their timing and temperature scores appear relatively higher only because these metrics are calculated on the small subset of PVs that matched correctly. 
This component-level breakdown is a practical diagnostic tool, indicating that evaluation should focus on core logical correctness, e.g. here the PV match.

\pgfplotstableread[col sep=comma]{%
model,label, fullscore, pvmatch, timing, temp
{claude-sonnet-4-20250514-thinking}, Claude-S4 (20250514), 90.73, 91.83, 85.55, 98.60
{claude-3.5-sonnet},                 Claude 3.5,           88.12, 88.51, 82.48, 97.59
{gpt-4o},                             GPT-4o,               85.04, 86.09, 80.63, 97.61
{gpt-5-high},                         GPT-5 High,           85.25, 85.74, 82.42, 96.97
{o3-high},                            o3-high,              84.71, 84.47, 80.61, 98.74
{grok-4-abacus},                      Grok-4 (abacus),      17.88, 17.20, 21.32, 62.20
{llama3.3},                           Llama 3.3,            49.50, 46.65, 65.23, 73.71
{qwen2.5},                            Qwen 2.5,             47.48, 41.35, 72.92, 72.79
}\datatable

\definecolor{oiBlue}{RGB}{0,114,178}        
\definecolor{oiOrange}{RGB}{230,159,0}      
\definecolor{oiBluishGreen}{RGB}{0,158,115} 
\definecolor{oiReddishPurple}{RGB}{204,121,167} 

\pgfplotstableread[col sep=comma]{
model,label,fullscore,pvmatch,timing,temp
{bs\_1},{Beamline Scientist 1},93.5,94.5,86.8,99.5
{claude-sonnet-4-20250514-thinking},{Claude Sonnet 4 (Thinking)},90.7,91.8,85.5,98.6
{gemini-2.5-pro-native},{Gemini 2.5 Pro},90.3,91.5,85.7,97.0
{gpt-5-high},{GPT-5 (high)},88.6,89.1,83.5,97.5
{claude-3.5-sonnet},{Claude 3.5 Sonnet},88.1,88.5,82.5,97.6
{grok-4-or},{Grok-4},87.1,88.1,80.4,97.5
{gpt-4o},{GPT-4o},85.0,86.1,80.6,97.6
{qwen3-coder-480b-or},{Qwen3-Coder (480B)},81.6,80.9,80.8,96.2
{llama3.3},{Llama3.3 (70.6B)},49.5,46.7,65.2,73.7
{qwen2.5},{Qwen2.5 (7.62B)},47.5,41.3,72.9,72.8
}\datatable

\begin{figure}[ht]
\makebox[\textwidth][c]{%
\begin{minipage}{1.10\textwidth}
\centering
\caption{EnvTrace full score and component metrics (PV match, timing, temperature) for complex-flow tasks. Temperature score is computed only when applicable.}
\label{fig:complex_tasks_components}

\begin{tikzpicture}
\begin{axis}[
    width=1.05\textwidth, height=8cm,
    ybar,
    bar width=5pt,
    enlarge x limits=0.06,
    symbolic x coords={ {Beamline Scientist 1},{Claude Sonnet 4 (Thinking)},{Gemini 2.5 Pro},{GPT-5 (high)},{Claude 3.5 Sonnet},{Grok-4},{GPT-4o},{Qwen3-Coder (480B)},{Llama3.3 (70.6B)},{Qwen2.5 (7.62B)} },
    xtick=data,
    xticklabel style={rotate=45,anchor=east,font=\scriptsize},
    ylabel={Complex Metrics (\%)},
    ymin=0, ymax=115,
    ymajorgrids=true,
    xmajorgrids=false,
    legend style={
        at={(0.5,1.02)},anchor=south,legend columns=-1,
        /tikz/every even column/.append style={column sep=0.6cm},
        draw=none, fill=none, font=\small
    },
    nodes near coords,
    every node near coord/.append style={font=\scriptsize,rotate=90,anchor=west,inner sep=0.8pt,text=black},
    nodes near coords filter/.code={\ifdim\pgfplotspointmeta pt=0pt\pgfkeyslet{/pgfplots/nodes near coords}{}\fi},
]

\addplot+[area legend, fill=oiBlue,         bar shift=-9pt] table[x=label,y=fullscore]{\datatable};
\addlegendentry{EnvTrace Full Score}

\addplot+[area legend, fill=oiOrange,       bar shift=-3pt] table[x=label,y=pvmatch]{\datatable};
\addlegendentry{PV Match}

\addplot+[area legend, fill=oiBluishGreen,  bar shift=+3pt] table[x=label,y=timing]{\datatable};
\addlegendentry{Timing}

\addplot+[area legend, fill=oiReddishPurple,bar shift=+9pt] table[x=label,y=temp]{\datatable};
\addlegendentry{Temp (applicable only)}

\end{axis}
\end{tikzpicture}

\vspace{0.3em}
\end{minipage}%
}
\end{figure}

\newpage

\section{Debug Baseline}

To establish an approximate upper bound for performance and to validate the stability of both the simulator and the EnvTrace framework, we conducted a series of \textit{debug} evaluations with the first code snippet from the GT.
%
Because the beamline control simulator is not fully deterministic due to small timing variations in device updates and asynchronous process scheduling, each debugging test was executed three times to account for the natural runtime variability of the environment. Timing variations also exist in real experiments.
As shown in Fig.~\ref{tab:debug_results}, even when the same ground-truth code is evaluated against itself, small variations in timing can lead to minor deviations in the final score. 
While identical executions would ideally yield a perfect score of 100\%, the simulator operates in real time with asynchronous device updates, introducing slight, expected timing jitter. 
These effects are small and consistent, serving mainly to illustrate that EnvTrace captures realistic execution variability rather than artificially idealized behavior. 
Accordingly, the debug scores represent an empirical upper bound of achievable performance under ideal functional conditions. 
Overall, the debug baselines demonstrate that EnvTrace is stable, reproducible within expected variability, and able to distinguish genuine functional correctness from superficial code similarity.

\begin{table}[H]
\centering
\caption{EnvTrace performance of debug ground-truth runs, representing an approximate upper bound of achievable scores.}
\label{tab:debug_results}
\begin{tabular}{lcc}
\toprule
\textbf{Evaluation Mode} & \textbf{EnvTrace Full Score (\%)} & \textbf{EnvTrace Accuracy (\%)} \\
\midrule
Simple-flow (Debug First GT)   & $98.1 \pm 0.6$ & $89.4 \pm 1.8$ \\
Complex-flow (Debug First GT)  &  $99.8 \pm 0.1$ & $100.0 \pm 0.0$ \\
\bottomrule
\end{tabular}
\end{table}

\section{Bluesky Experiments}

\begin{figure}[ht]
\makebox[\textwidth][c]{%
\begin{minipage}{1.12\textwidth}
\centering
\caption{Model performance on the generic Bluesky-plan dataset (N=20). Left bars show the EnvTrace full score and right stacked bars give the string-based exact match (bottom, orange) and improvement (top, green) with EnvTrace accuracy.}
\label{fig:bluesky_performance}

\begin{tikzpicture}
\begin{axis}[
    width=1.05\textwidth, height=8cm,
    ybar,
    bar width=4pt,
    enlarge x limits=0.06,
   symbolic x coords={ {o3 (high)},{GPT-4o (Abacus)},{Claude Opus 4 (Thinking)},{Claude Sonnet 4 (Thinking)},{Claude 3.5 Sonnet},{GPT-4o},{Qwen2.5-Coder (32.8B)},{Llama3.3 (70.6B)},{Athene-v2 (72.7B)},{Mistral-NeMo (12.2B)},{Phi-3.5 (3.8B)},{Qwen3 (32B)} },
    xtick=data,
    xticklabel style={rotate=45,anchor=east,font=\scriptsize},
    ylabel={Performance (\%)},
    ymin=0, ymax=115,
    ymajorgrids=true,
    xmajorgrids=false,
    legend style={
        at={(0.5,1.02)}, anchor=south,
        legend columns=-1, /tikz/every even column/.append style={column sep=0.6cm},
        draw=none, fill=none, font=\small
    },
]

\addplot+[area legend, fill=oiBlue, bar shift=-2.5pt,
          nodes near coords,
          every node near coord/.append style={font=\scriptsize,rotate=90,anchor=west,inner sep=1pt,text=black},
          point meta=y]
    table[x=label,y=fullscore]{\datatable};
\addlegendentry{EnvTrace Full Score (\%)}

\addplot+[area legend, ybar, stack plots=y, fill=oiOrange, bar shift=+2.5pt]
    table[x=label,y=exact]{\datatable};
\addlegendentry{Exact Match (\%)}

\addplot+[area legend, ybar, stack plots=y, fill=oiBluishGreen, bar shift=+2.5pt, forget plot]
    table[x=label,y=improvement]{\datatable};

\addlegendimage{area legend, ybar, draw=none, shade,
  bottom color=oiOrange, top color=oiBluishGreen}
\addlegendentry{EnvTrace Accuracy (\%)}

\addplot+[draw=none, fill=none, bar shift=+2.5pt,
          nodes near coords, point meta=explicit,
          every node near coord/.append style={
            font=\scriptsize, rotate=90, anchor=west, inner sep=1pt, text=black
          }]
    table[x=label,y=total,meta expr=\thisrow{total}]{\datatable};

\end{axis}
\end{tikzpicture}
\end{minipage}%
}
\end{figure}
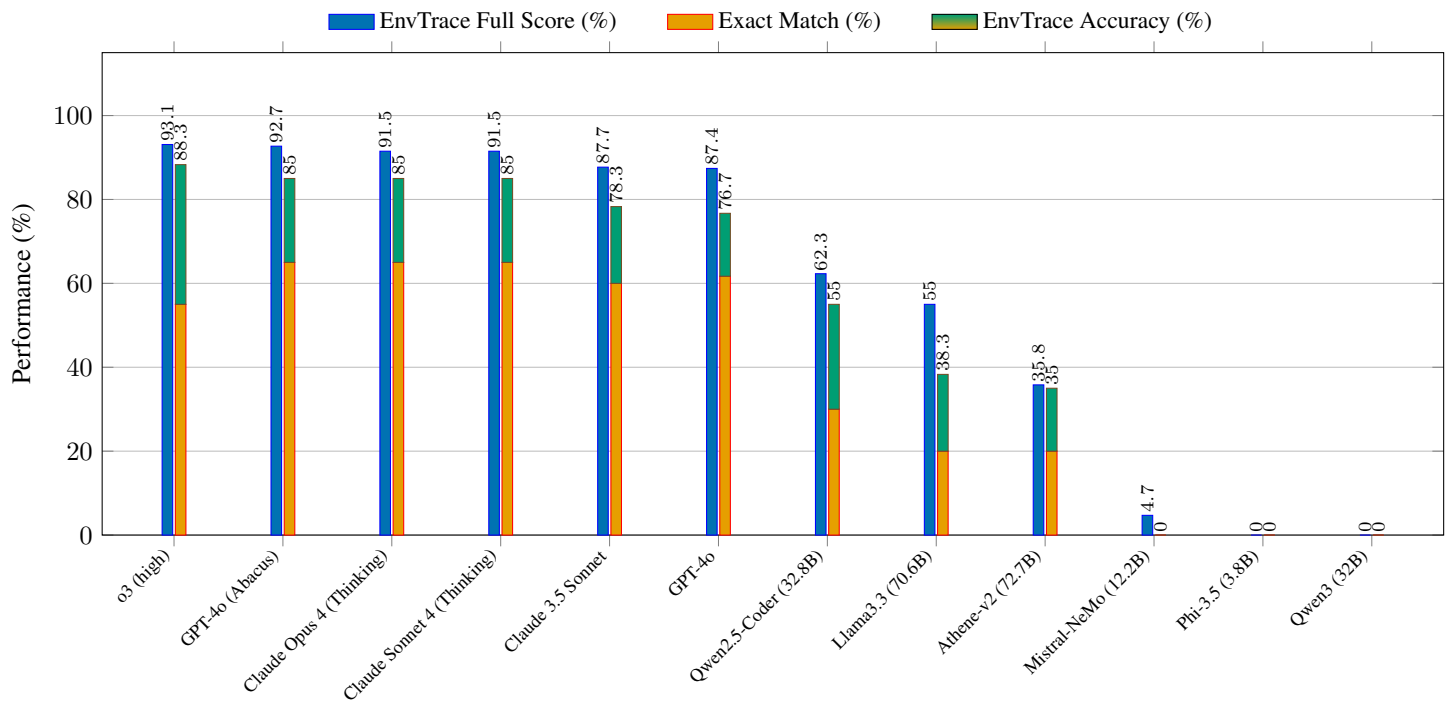

To show the generalizability of VISION~\cite{mathur2024vision} and EnvTrace, we have adapted the operator cog (the part of the tool that takes care of generating operations) to produce code for a small benchmark dataset of generic Bluesky plans. 
These generic Bluesky plans can be applied across multiple Bluesky-supported beamlines without requiring additional documentation or code changes but only function input parameters need to be specified. A \textit{beamline specifics} text file is dynamically loaded and included into the system prompt. This file contains beamline specific information, such as detector or motor names.
These Bluesky plan definitions were parsed from the Bluesky documentation. Usage examples were produced both manually and through LLM generation with documentation reference and human verification.
LLM performance for generic Bluesky plans is shown in Table~\ref{tab:simple_bluesky_results} and Fig.~\ref{fig:bluesky_performance}.

\begin{table}[h]
  \centering
  \begin{threeparttable}
    \caption{Performance of LLMs on generic Bluesky tasks (N=20)}
    \label{tab:simple_bluesky_results}
    \sisetup{separate-uncertainty=true, table-align-uncertainty=true}
    \begin{tabular}{
      l
      S[table-format=2.2(2.2)]
      S[table-format=2.2(2.2)]
      S[table-format=2.2(2.2)]
      S[table-format=2.2(2.2)]
      S[table-format=2.2(2.2)]
    }
      \toprule
      \textbf{Model} & 
      {\textbf{\begin{tabular}[c]{@{}c@{}}EnvTrace \\ Full Score (\%)\tnote{a}\end{tabular}}} & 

      {\textbf{\begin{tabular}[c]{@{}c@{}} EnvTrace \\ Accuracy (\%)\tnote{b}\end{tabular}}} & \textcolor{gray}
      {\textbf{\begin{tabular}[c]{@{}c@{}}Exact Match \\ Accuracy (\%)\tnote{c}\end{tabular}}} & \textcolor{gray}
      {\textbf{\begin{tabular}[c]{@{}c@{}}Norm. Lev. \\ Dist. (\%)\tnote{d}\end{tabular}}} &
      {\textbf{\begin{tabular}[c]{@{}c@{}}Inference \\ Time (s)\end{tabular}}} \\
      \midrule
\multicolumn{6}{l}{\textbf{Closed Source Models}} \\
\midrule
\multicolumn{6}{l}{\textit{OpenAI / Microsoft}} \\
  o3 (high)\tnote{t} & 93.1 \pm 2.4 & 88.3 \pm 2.9 & 55.0 \pm 0.0 & 11.5 \pm 1.3 & 4.6 \pm 0.8 \\
  GPT-4o (Abacus) & 92.7 \pm 0.2 & 85.0 \pm 0.0 & 65.0 \pm 0.0 & 7.2 \pm 0.0 & 1.9 \pm 0.1 \\
  GPT-4o & 87.4 \pm 4.5 & 76.7 \pm 2.9 & 61.7 \pm 2.9 & 7.4 \pm 0.2 & 1.5 \pm 0.2 \\
\midrule
\multicolumn{6}{l}{\textit{Anthropic}} \\
  Claude Opus 4 (Thinking)\tnote{t} & 91.5 \pm 0.1 & 85.0 \pm 0.0 & 65.0 \pm 0.0 & 7.2 \pm 0.0 & 8.1 \pm 0.3 \\
  Claude Sonnet 4 (Thinking)\tnote{t} & 91.5 \pm 0.0 & 85.0 \pm 0.0 & 65.0 \pm 0.0 & 7.2 \pm 0.0 & 5.9 \pm 0.2 \\
  Claude 3.5 Sonnet & 87.7 \pm 0.1 & 78.3 \pm 2.9 & 60.0 \pm 0.0 & 7.6 \pm 0.0 & 2.9 \pm 0.2 \\
\midrule
\multicolumn{6}{l}{\textbf{Open Source Models}} \\
\midrule
  Qwen2.5-Coder (32.8B) & 62.3 \pm 0.0 & 55.0 \pm 0.0 & 30.0 \pm 0.0 & 26.5 \pm 0.0 & 3.4 \pm 0.0 \\
  Llama3.3 (70.6B) & 55.0 \pm 0.2 & 38.3 \pm 2.9 & 20.0 \pm 0.0 & 29.9 \pm 0.0 & 7.1 \pm 0.0 \\
  Athene-v2 (72.7B) & 35.8 \pm 0.1 & 35.0 \pm 0.0 & 20.0 \pm 0.0 & 19.2 \pm 0.0 & 6.4 \pm 0.0 \\
  Mistral-NeMo (12.2B) & 4.7 \pm 0.0 & 0.0 \pm 0.0 & 0.0 \pm 0.0 & 34.5 \pm 0.0 & 1.3 \pm 0.0 \\
  Phi-3.5 (3.8B) & 0.0 \pm 0.0 & 0.0 \pm 0.0 & 0.0 \pm 0.0 & 84.3 \pm 1.8 & 5.5 \pm 1.2 \\
  Qwen3 (32B) & 0.0 \pm 0.0 & 0.0 \pm 0.0 & 0.0 \pm 0.0 & 96.8 \pm 0.0 & 48.7 \pm 0.0 \\
\bottomrule
    \end{tabular}
    \begin{tablenotes}
      \small
      \item[a] \textbf{EnvTrace Full Score (\%):} The continuous score from 0 to 100, reflecting the weighted average of state and temporal fidelity.
      \item[b] \textbf{EnvTrace Accuracy (\%):} The percentage of test cases passing the strict binary criteria for semantic equivalence.\item[c] \textbf{Exact Match (\%):} The percentage of test cases where the generated code string is identical to a ground-truth solution \cite{mathur2024vision}.
        \item[d] \textbf{Normalized Levenshtein Distance (\%):} A value of $0\%$ indicates identical strings.
      \item[e] All results are the mean and standard deviation over three runs. \item[f] Models are grouped and then sorted by descending Average Full Score.
      \item[t] Models run with reasoning enabled. ``(high)'' and ``(minimal)'' denote different reasoning levels.
    \end{tablenotes}
  \end{threeparttable}
\end{table}

\pgfplotstableread[col sep=semicolon]{
label;fullscore;exact;improvement;total
{o3 (high)};93.1;55.0;33.3;88.3
{GPT-4o (Abacus)};92.7;65.0;20.0;85.0
{Claude Opus 4 (Thinking)};91.5;65.0;20.0;85.0
{Claude Sonnet 4 (Thinking)};91.5;65.0;20.0;85.0
{Claude 3.5 Sonnet};87.7;60.0;18.3;78.3
{GPT-4o};87.4;61.7;15.0;76.7
{Qwen2.5-Coder (32.8B)};62.3;30.0;25.0;55.0
{Llama3.3 (70.6B)};55.0;20.0;18.3;38.3
{Athene-v2 (72.7B)};35.8;20.0;15.0;35.0
{Mistral-NeMo (12.2B)};4.7;0.0;0.0;0.0
{Phi-3.5 (3.8B)};0.0;0.0;0.0;0.0
{Qwen3 (32B)};0.0;0.0;0.0;0.0
}\datatable

\section{Semantic versus Syntactic Metrics}
\label{si:relationship_analysis}

To further support the claim that syntactic metrics are an unsuitable proxy for functional correctness, here we show the semantic EnvTrace full score against CodeBLEU (comb 7) and normalized Levenshtein distance (nLD) for simple- and complex-flow tasks based on Table~\ref{tab:simple_results} and Table~\ref{tab:complex_results}, respectively.

\begin{figure}[ht!]
    \centering
    \includegraphics[width=0.8\textwidth]{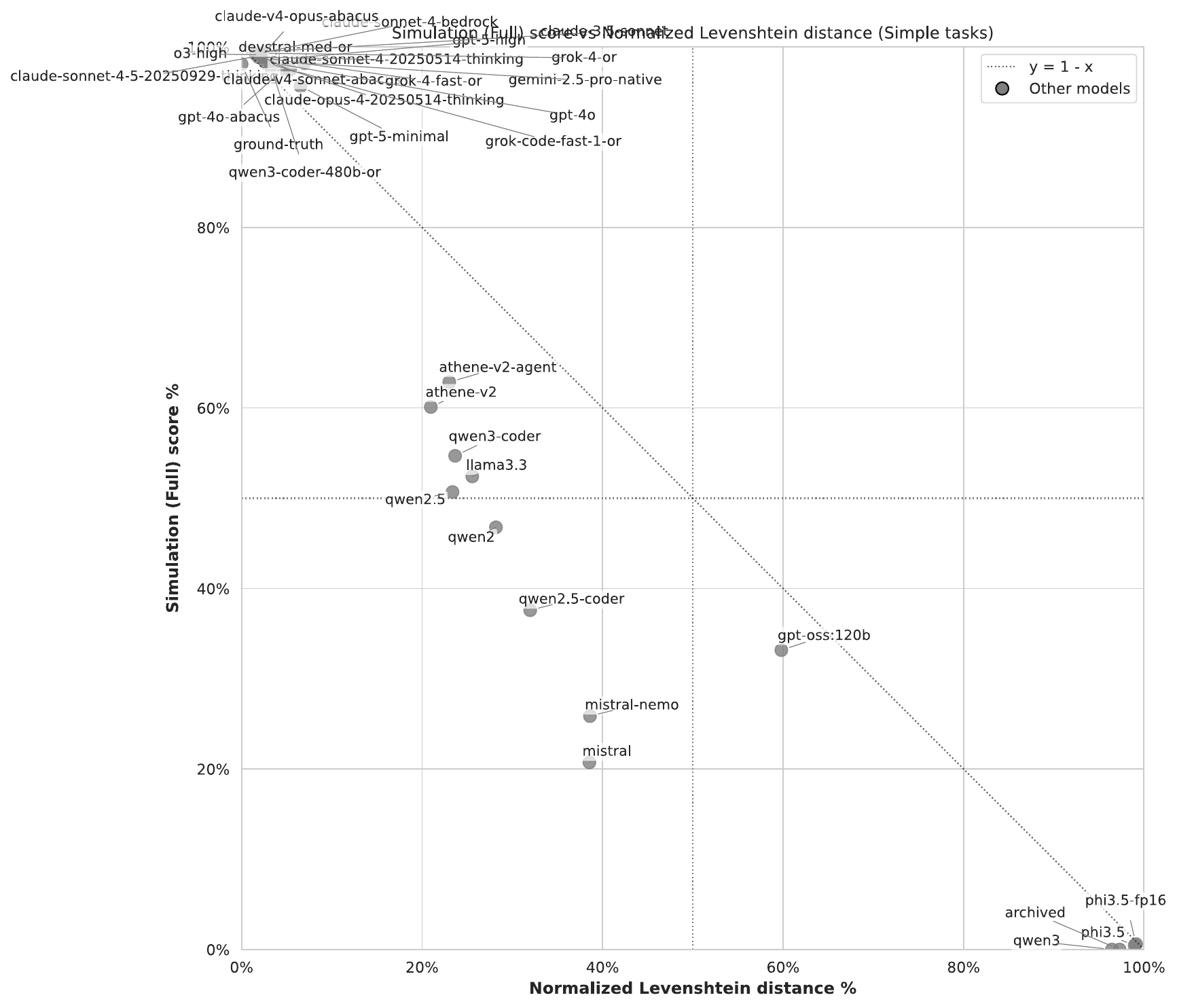}
    \caption{Plot of the semantic EnvTrace full score versus the normalized Levenshtein distance for simple-flow tasks.}
    \label{fig:si_levenshtein_simple}
\end{figure}

\begin{figure}[ht!]
    \centering
    \includegraphics[width=0.8\textwidth]{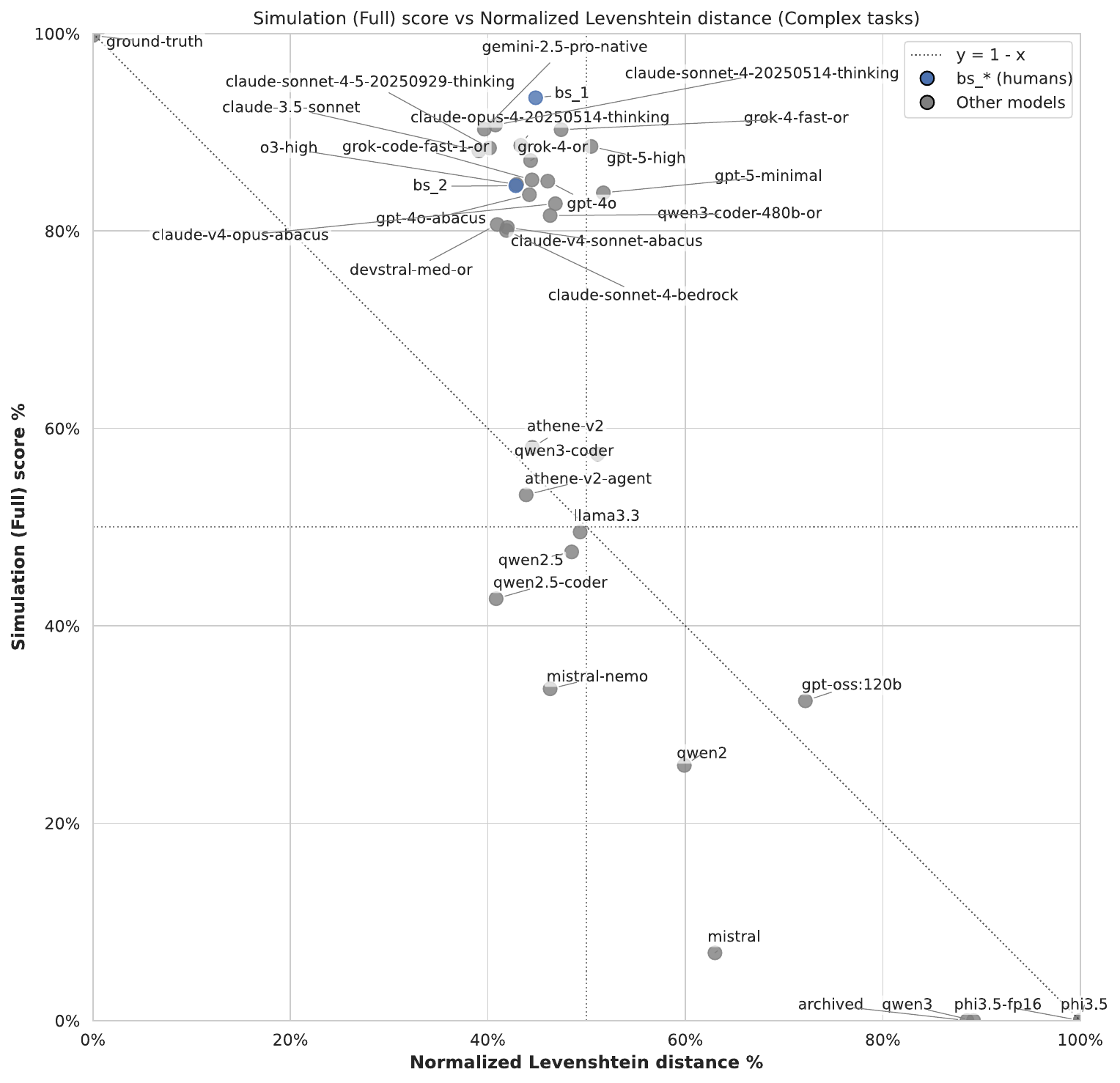}
    \caption{Plot of the semantic EnvTrace full score versus the normalized Levenshtein distance for complex-flow tasks. }
    \label{fig:si_levenshtein_complex}
\end{figure}

\begin{figure}[ht!]
    \centering
    \includegraphics[width=0.8\textwidth]{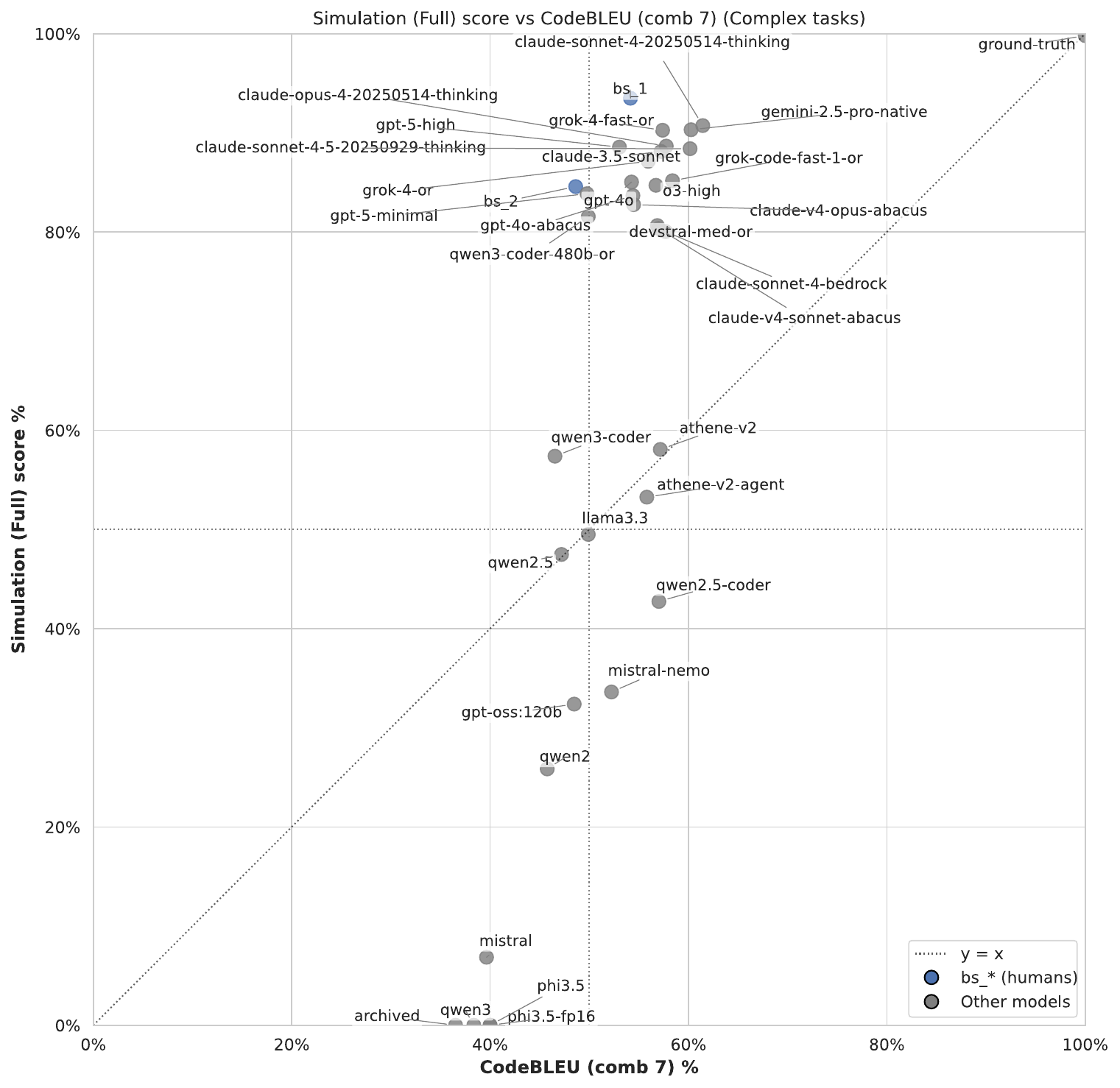}
    \caption{Plot of the semantic EnvTrace full score versus the syntactic CodeBLEU score for complex-flow tasks.}
    \label{fig:si_codebleu_complex}
\end{figure}

For simple-flow tasks consisting of single or short sequences of commands, the relationship between EnvTrace full score and nLD is shown in Fig.~\ref{fig:si_levenshtein_simple}. 
The CodeBLEU metric is not applicable for simple-flow tasks, as data-flow analysis component requires more complex code structures to function correctly; for single-line or short commands, there is typically no data-flow to analyze.
Complex flow tasks are those that require the generation of structured control logic, such as \texttt{for} or \texttt{while} loops, and conditional \texttt{if} statements. For these tasks, the potential for stylistic and logical variation is high, making them suitable tests for the robustness of evaluation metrics. Figure~\ref{fig:si_codebleu_complex} shows the relationship between EnvTrace score and CodeBLEU, and Fig.~\ref{fig:si_levenshtein_complex} for nLD. 
The dotted diagonal lines represent a hypothetical perfect one-to-one mapping between the syntactic and semantic scores (i.e., \texttt{y = x} for CodeBLEU and \texttt{y = 1 - x} for Levenshtein distance). The distance of a data point from the ideal line reflects the extent to which the syntactic metric may be unreliable.

While syntactic metrics can be useful for "sanity-checking" simple code generation (Fig.~\ref{fig:si_levenshtein_simple}), their utility rapidly diminishes as task complexity increases (Fig.~\ref{fig:si_levenshtein_complex}), reinforcing the need for a robust semantic evaluation framework.
As seen in Figure~\ref{fig:si_codebleu_complex}, many models lie far from the ideal line. Points located significantly above the line, such as \texttt{Slaude-3.5-Sonnet} and \texttt{Claude-Opus-4- (Thinking)}, represent "false negatives" for the CodeBLEU metric; their functional performance is much higher than their syntactic similarity score would suggest.
Conversely, points below the line, such as \texttt{Athene-v2} and \texttt{Qwen2.5-Coder}, are "false positives"; their syntactic similarity is misleadingly high given their lower functional performance. 

\section{AI Evaluator}
\label{si:ai}
\begin{figure}[H]
  \centering
  \includegraphics[width=\textwidth]{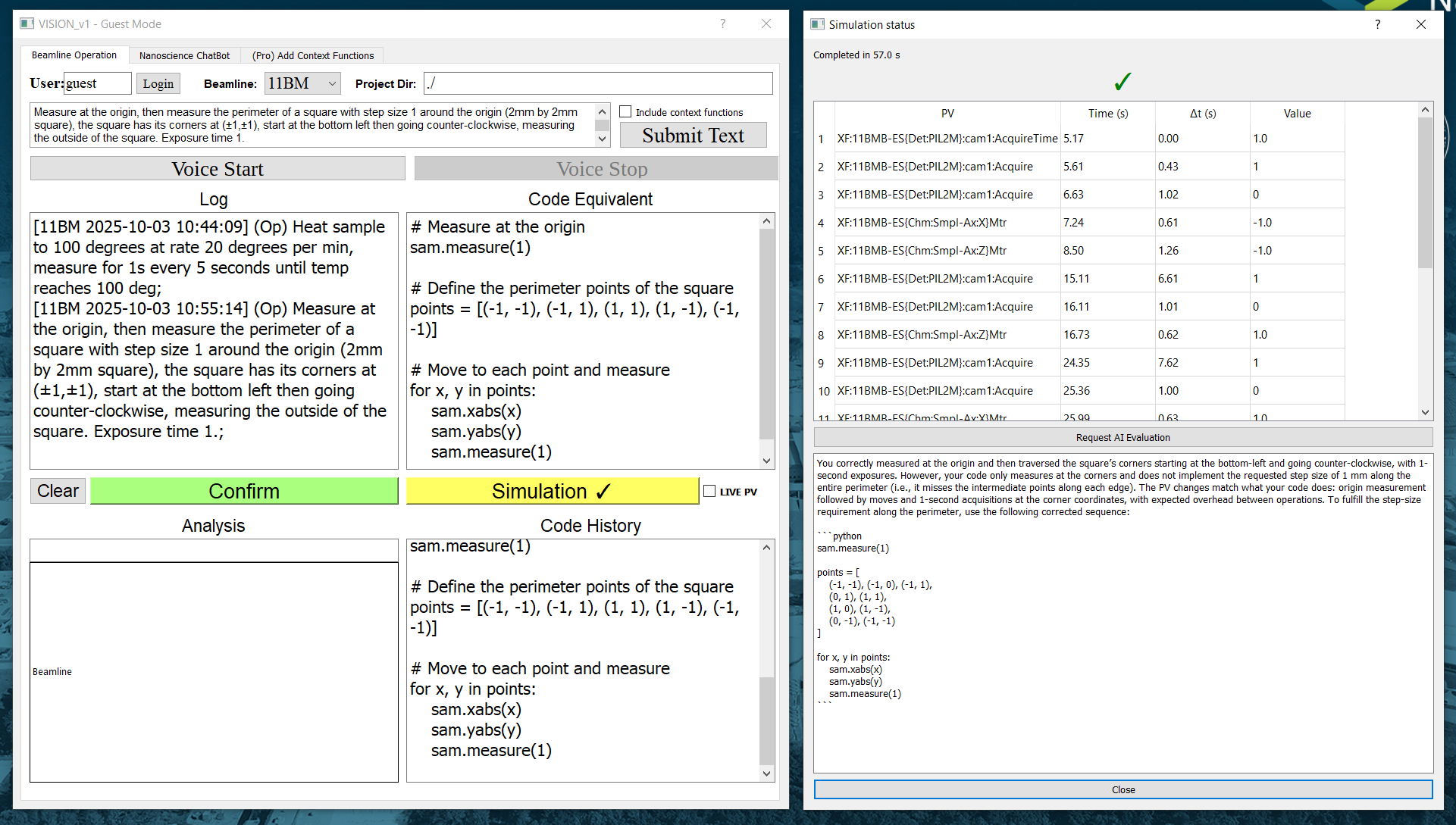}
  \caption{Example of the AI evaluator providing corrections.}
  \label{si:fig:square_example}
\end{figure}

In the VISION GUI, once a user has completed a simulation of the generated code, they can request an AI evaluation. This AI evaluation is driven by a prompt that takes in the user's natural language query, the results of the simulation (the tracked PV changes), the operator cog prompt (the dynamically generated, beamline specific, coding prompt), and the generated code. Once the AI evaluation button is passed, it will give a short evaluation commenting on the pipeline (natural language query to code to PV trace) to the user.
Here we used \texttt{GPT-5-low} as the AI evaluator, as it strikes a good balance between inference time and performance.
An example in which AI evaluation provides correction is given in Fig.~\ref{si:fig:square_example}.

\end{document}